# Tackling the dimensions in imaging genetics with CLUB-PLS

Andre Altmann[1], Ana C Lawry Aguila[1], Neda Jahanshad[2], Paul M Thompson[2], Marco Lorenzi[3]

Affiliations

1 Centre for Medical Image Computing (CMIC), Department of Medical Physics and Biomedical Engineering, University College London (UCL), London, UK

2 Imaging Genetics Center, Stevens Neuroimaging and Informatics Institute, Keck School of Medicine of USC, Marina del Rey, CA, USA

3 Université Côte d'Azur, Inria Sophia Antipolis, Epione Research Project, France

Corresponding author:

Andre Altmann

Centre for Medical Image Computing (CMIC)

Department of Medical Physics and Biomedical Engineering

University College London

90 High Holborn, 1st Floor

London WC1V 6LJ

Email: a.altmann@ucl.ac.uk

**Abstract**

A major challenge in imaging genetics and similar fields is to link high-dimensional data in one domain, e.g., genetic data, to high dimensional data in a second domain, e.g., brain imaging data. The standard approach in the area are mass univariate analyses across genetic factors and imaging phenotypes. That entails executing one genome-wide association study (GWAS) for each pre-defined imaging measure. Although this approach has been tremendously successful, one shortcoming is that phenotypes must be pre-defined. Consequently, effects that are not confined to pre-selected regions of interest or that reflect larger brain-wide patterns can easily be missed. In this work we introduce a Partial Least Squares (PLS)-based framework, which we term Cluster-Bootstrap PLS (CLUB-PLS), that can work with large input dimensions in both domains as well as with large sample sizes. One key factor of the framework is to use cluster bootstrap to provide robust statistics for single input features in both domains. We applied CLUB-PLS to investigating the genetic basis of surface area and cortical thickness in a sample of 33,000 subjects from the UK Biobank. We found 107 genome-wide significant locus-phenotype pairs that are linked to 386 different genes. We found that a vast majority of these loci could be technically validated at a high rate: using classic GWAS or Genome-Wide Inferred Statistics (GWIS) we found that 85 locus-phenotype pairs exceeded the genome-wide suggestive ($P<1e-05$) threshold.

## Introduction

Imaging genetics strives to identify genetic loci that influence imaging-derived phenotypes. Thus far, the imaging genetics approach is most widely applied in the neuroscience field where the phenotypes are typically obtained from magnetic resonance imaging (MRI) scans of the human brain. One appeal of the imaging genetics approach, in contrast to classic case-control genetic analyses, is that phenotypes derived from imaging data can quantify ongoing disease processes and thus boost statistical power to identify disease relevant genetic loci. Moreover, the imaging genetics approach can be used to identify genetic influences on brain development and brain aging that are independent of disease.

Since its conception, the prevailing method in imaging genetics remains the mass univariate genome-wide association study (GWAS) approach where each imaging phenotype is tested individually against all SNPs using linear regression. Growing sample sizes and world-wide collaborative studies (Medland *et al.*, 2014) have led to a great success of this approach in identifying genetic influences on intracranial volume (Stein *et al.*, 2012), subcortical volumes (Hibar *et al.*, 2015), cortical morphology (Grasby *et al.*, 2020), and many other MRI-derived phenotypes (Elliott *et al.*, 2018). However, the application of the standard GWAS methodology has limitations. The main one being that imaging phenotypes have to be preselected and studies are therefore often based on available brain parcellations (Stein *et al.*, 2012; Medland *et al.*, 2014; Hibar *et al.*, 2015; Elliott *et al.*, 2018; Grasby *et al.*, 2020). Genetic effects that influence sub-regions or cross regional boundaries may therefore evade detection. Moreover, influences on large-scale brain organization, such as left-right asymmetry or expansion along the anterior-posterior gradient, may be missed entirely because they have not been explicitly modeled.

Numerous machine learning approaches are being investigated to overcome the limitation of mass-univariate testing in imaging genetics (Shen and Thompson, 2019). However, learning-based approaches face a major challenge due to the low explanatory power of individual single nucleotide polymorphisms (SNPs), the main genetic marker investigated in imaging genetics, on phenotypes. In recent work, Shin *et al.* (2020) first applied principal component analysis (PCA) to the imaging data and then investigated the genetic associations with the top two principal components (PCs) using a standard GWAS approach. In a related method, we previously used partial least squares (PLS) for an integrated analysis of genome-wide and vertex-wide data and identified genetic loci modulating the typical pattern of Alzheimer's brain pathology (Lorenzi *et* *al.*, 2018), however, the memory constraints of this approach prohibits its application to large datasets such as the UK Biobank (UKB).

There exist further methods in the statistical genetics field that capture the genetic effects of single genetic variants on complex multivariate phenotypes. For instance, canonical correlation analysis (CCA)-based approaches extract the linear combination of phenotypes that explain the largest possible amount of the covariation between the SNP and all phenotypes (Ferreira and Purcell, 2009). Effectively, this test discovers whether there is an association between the SNP and any of the phenotypes. The concept has been used to investigate the genetic architecture of face and brain morphology (Claes *et al.*, 2018; Naqvi *et al.*, 2021). A further recent example is the Multivariate Omnibus Statistical Test (MOSTest) (van der Meer *et al.*, 2020), which leverages shared genetic signal between phenotypes to boost statistical power. These approaches attempt to partially alleviate the multiple-testing burden in imaging genetics application by exploiting the correlation structure of the imaging phenotypes. However, these methods treat every genetic variant independently of the rest and essentially remain standard massively univariate GWAS. Notably, the imaging genetics field has generated a range of machine learning approaches (Shen and Thompson, 2019). For instance, sparse reduced rank regression, which is a special type of multivariate multiple regression models for identifying multi-SNP and multi-phenotype associations (Vounou *et al.*, 2012). Other approaches in this area seek to integrate knowledge about gene-gene interactions and other biological priors such as molecular pathways as constrains in structured regularized machine learning approaches, e.g., (Silver *et al.*, 2012). However, despite their appeal, often the computational burden for these approaches requires pre-selection of a few thousand SNPs. Thus, rendering them ill-suited for discovery analyses.

Here we extend our previous work and introduce CLUster Bootstrap PLS (CLUB-PLS), which enables computing PLS solutions for high dimensional data as well as arbitrary large sample sizes and provides at the same time a statistical interpretation of feature weights using an efficient bootstrap approach. CLUB-PLS is applied to 33,000 participants of the UK Biobank project.

## Methods

*Discovery*

The discovery data comprises N=38,724 subjects from UK Biobank (UKB) with available genome-wide genotyping data and T1-weighted MRI imaging data. We restricted the subjects to central European ancestry based on genetic clustering provided by UKB and successful extraction of regional surface area (66 regions) and regional cortical thickness (66 regioins) based on the Desikan-Killiany (DK) atlas (Desikan *et al.*, 2006) obtained using FreeSurfer v6.0 (Dale *et al.*, 1999; Fischl *et al.*, 1999). This resulted in a set of 33,725 eligible subjects for the analysis.

The number of genotyped SNPs was 805,426. For the analysis we retained only autosomal SNPs with a minor allele frequency of at least 1% and a maximum missingness rate of 2% leading to a set of 582,538 SNPs. Following these steps, 635 subjects with a high fraction of missing SNPs were excluded, resulting in a final set of 33,070 subjects. We used the conventional additive coding for SNPs, i.e., counting the number of minor alleles. The proposed analysis approach requires a complete genetic matrix without missing values; thus we used mean imputation to fill the few missing values. Furthermore, each SNP was centered (i.e., the mean was zero).

The imaging data comprising 132 regions are subject to influences from age, sex and intra cranial volume (ICV). Thus, we regressed the effect of age (at imaging), age$^2$, ICV, sex and the first 20 genetic principal components for population structure using linear regression. The mean of the resulting residuals was zero and, in addition, the residuals were scaled to standard deviation of 1.0.

*Partial Least Squares*

The statistical approach for this imaging genetic analysis is based on the partial least squares (PLS) method. PLS aims to find linear projections for the two input modalities such that in the projected space they have maximal covariation. In particular, we relied on the PLS version implemented using singular value decomposition (SVD). Briefly, let $X \in R^{Nxp1}$ be the matrix of imaging data (p1=number of imaging features), $Y \in R^{Nxp2}$ be the matrix of genetic data (p2=number of genetic features), and $N$ be the sample size. Then the cross-covariance matrix $C = X^T Y$ can be decomposed using SVD:

$$C = U\,D\,V^T$$

, where the columns of $U$ are the left singular vectors of $C$, the columns of $V$ are the right singular vectors and $D$ is a diagonal matrix of the singular values. In particular, $U$ contains the weights for the imaging features, while $V$ contains the weights for the genetic features such that the projections of $U$ and $V$ onto the data matrices $X$ and $Y$ maximally covary.

However, in cases where *p1* and *p2* are high-dimensional, the cross-covariance matrix becomes prohibitively large. Previously, we used the approach described by Worsley *et al.* (2005) based on eigenvalue decomposition that avoids the explicit computation of $C$ to conduct an imaging genetics analysis on SNP-level and vertex-level (Lorenzi *et al.*, 2018):

$$(X^T X Y^T Y)A = AL$$

, where the columns of $A$ contain the eigenvectors and $L$ contains the eigenvalues, with $D = L^{1/2}$. Setting $B = A(A^T Y^T Y A)^{-1/2}$. Then, the left singular matrix $U$ can be then computed as $U = X(Y^T Y B L^{-1/2})$, while the right singular matrix $V$ can be computed as $V = YB$. Thus, avoiding the need to calculate the *p1* by *p2* matrix $C$. However, this does not alleviate the computational challenges when $N$ is large.

We recently extended this PLS approach to afford computation in a federated fashion where data are distributed across different sites (Lorenzi *et al.*, 2017). Briefly, in this federated or *meta PLS* setting each site computes a local solution to the PLS problem using only data that is available at that site, then the local solutions are shared with a central processor that computes the approximated global solution reflecting all participating samples. In this work, we employed meta PLS to scale the PLS approach to work with datasets comprising many subjects (i.e., large *N*). By splitting a locally available large dataset into several (equally sized) 'chunks', this approach enables us to work locally with arbitrary large dataset: the PLS solution for each chunk is computed and then combined in the meta PLS fashion to obtain the approximated overall solution, similar to the group-PCA approach used in fMRI analyses (Smith *et al.*, 2014). Here, we randomly selected 33,000 subjects and split the available data into 66 chunks each comprising 500 subjects. The combination of meta PLS with the eigenvalue approach enables us to obtain PLS solutions for views with high dimensionality (large *p1* and/or *p2*) as well as arbitrary large sample sizes (large *N*).

*Cluster bootstrap*

A single execution of the meta PLS algorithm using all 66 clusters results in the matrices $U$ and $V$, containing weights for the imaging and genetic PLS components, respectively. However, this single execution lacks information on the variability of these PLS weights and therefore does not provide the means to assess whether the weights are in fact substantially different from 0; thus *p*-values for PLS weights cannot be obtained. In previous work (Lorenzi *et al.*, 2018), we used repeated split-half cross-validation to estimate the stability selection of PLS weights and thereby their feature importance. In the absence of a sound theoretical framework to compute standard errors of the PLS weights, we propose using the bootstrap (Efron, 1992) as a wide-spread empirical method to obtain such standard error estimates. However, a straightforward application of the bootstrap would entail sampling with replacement from the entire dataset comprising 33,000 subjects and repeating the entire pipeline each time. Thus, creating a significant computational overhead. Instead, we apply the 'cluster bootstrap', where the sampling is not applied to the subject-level data, but to pre-defined sets or clusters. In our application, we use as clusters the 66 chunks comprising 500 subjects each. The cluster bootstrap has been shown to work with a sufficiently large number of clusters (30 or more) (Cameron *et al.*, 2008). By combining the cluster bootstrap with meta PLS we can leverage the pre-computed PLS solutions for each cluster and compute the desired standard errors for PLS weights in a computationally efficient way. We computed 500 bootstrap replications of the PLS on the entire set of 33,000 subjects to obtain means and standard deviations of PLS weights and to compute *p*-values based on the Wald statistic. Briefly, each of the *k=1,...,K* cluster bootstrap executions results in a left and right singular matrices ${U^*}_k$ and ${V^*}_k$, respectively. (We will show the remaining derivation only for $U$ since the same applies to $V$.) Next, we can estimate the average bootstrap estimate $\bar{U}^* = \frac{1}{K}\sum_{k=1}^{K} {U^*}_k$. Likewise, we can estimate the bootstrap standard error $S^* = ( \frac{1}{K-1} \sum_{k=1}^{K} (\bar{U}^* - {U^*}_k)^2 )^{1/2}$.

Now, for feature $i \in \{1,\ldots,p1\}$ in PLS component $c$ we get following Wald statistic: $w_{i,c} = (U_{i,c} - \bar{U}_{i,c}^{*})/{S^*}_{i,c}$, which is asymptotically normally distributed if the number of clusters is large enough. Thus, using cluster bootstrap we can derive *p*-values for imaging features and genetic features from the PLS method. Moreover, as a benefit of using the (cluster) bootstrap there are (clusters of) subjects that have not been used for estimating the PLS model and therefore these subjects can be used to estimate the out-of-sample correlations between the imaging and genetics projections. We refer to this combination of cluster bootstrap and meta PLS as CLUB-PLS (for CLUster Bootstrap PLS).

*Application of CLUB-PLS to the UK Biobank*

To illustrate the potential of CLUB-PLS, we computed the first ten PLS components for cortical thickness (CT) and surface area (SA) each paired with genome-wide SNP data. Within each genetic component we set the genome-wide significance threshold of P<5e-08 to identify loci of interest. Genome-wide significant loci were annotated using the SNP2GENE tool in FUMA (v1.5.5) (Watanabe *et al.*, 2017). SNPs were mapped to genes by position (within 10kb distance), eQTL (using all brain tissues available in GTEx v8 (GTEx Consortium, 2015)), and 3D chromosome interactions maps (fetal and adult cortex (Giusti-Rodr\'\iguez *et al.*, 2018) , hippocampus and neural progenitor cells (Schmitt *et al.*, 2016)).

*Validation using UK Biobank GWAS summary statistics*

We assessed how well the discovered genetic loci using CLUB-PLS can be replicated using traditional GWAS approaches. To this end we relied on results from mass univariate testing for individual regions' CT and SA measures carried out in the UK Biobank (Smith *et al.*, 2021). We used the GWIS method (Nieuwboer *et al.*, 2016) to convert the associations with individual regions to statistics for multivariate imaging patterns generated by CLUB-PLS. In brief, we used the projection weights for the imaging component $U$ to provide the weights for the linear combination of existing summary statistics, similar to the work by Shin *et al.* (2020). We expect highly significant loci to be identified by CLUB-PLS to also exhibit a strong signal in the GWIS-based validation.

## Results

We have applied our CLUB-PLS approach to investigate the link between genetics and cortical surface area as well as cortical thickness in 33,000 subjects of the UK Biobank. The bootstrap component enabled us to compute out-of-sample correlation results between the genetic projection and the projection of the imaging data (**Table 1**). Among the first ten PLS components, correlations between surface area (SA) and genetics were higher (Pearson's r ranging from 0.084 to 0.184) than between cortical thickness (CT) and genetics (Pearson's r ranging from 0.006 to 0.118). There were 79 and 28 loci exceeding the cutoff for genome wide significance (P=5e-08) across the top five PLS components for SA (**Table 2**) and CT (**Table 3**), respectively.

*Genetics of cortical surface area*

**Figure 1** depicts the Manhattan plots for the first five components along with the associated brain maps. There are 18 loci exceeding the genome-wide significance threshold (P<5e-8) for Component 1. The imaging phenotype associated with this component models overall brain surface area (as indicated by all imaging weights having the same sign; **Figure 1a**). The imaging phenotype associated with component 2 scales the surface area of the occipital lobe (**Figure 1b**) and there are 40 associated loci. Component 3 describes reduced area in the inferior parietal, the inferior and middle temporal lobes as well as the banks of the superior and temporal sulcus. At the same time the component describes an increase in area of the transverse temporal and the superior frontal lobes (**Figure 1c**). The matching genetic component yields 18 significant loci. Component 4 models reduction in the postcentral gyrus and paracentral lobule and increase frontal pole, middle temporal gyrus and pars orbitalis and was only associated with a single genome-wide significant locus (**Figure 1d**). Lastly, component 5 describes decreases in SA for superior temporal gyrus and transverse temporal gyrus and SA increases in lateral orbitofrontal cortex, pars orbitalis and medial orbitofrontal cortex. There were two genome-wide significant loci associated with this pattern (**Figure 1e**).

*Genetics of cortical thickness*

**Figure 2** depicts the Manhattan plots for the first five components along with the associated brain maps. There are 17 loci exceeding the genome-wide significance threshold (P<5e-8) for PLS Component 1. Like in the case of Surface Area, the imaging phenotype associated with this component models overall cortical thickness (as indicated by all imaging weights having the same sign; **Figure 2a**). The imaging phenotype associated with component 2 models brain-wide asymmetry in cortical thickness between the right and the left hemisphere (**Figure 2b**). However, no loci reached genome-wide significance for this pattern. CT PLS component 3 models reduced cortical thickness in the insula cortex, the medial orbitofrontal cortex and the superior temporal gyrus. Conversely, CT in the precuneus as well as the superior parietal lobule are increased (**Figure 2c**). There are six independent loci significantly associated with this pattern. The CT PLS component 4 models increased CT of the superior frontal gyrus, the caudal and rostal middle frontal gyri as well as decreased CT of the lingual and the parahippocampal gyri (**Figure 2d**). There are four genome-wide significant loci associated with this pattern. Only one significant loci was associated with the pattern described by CT PLS component 5: increased cortical thickness along the cingulate and decreased cortical thickness entorhinal cortex, the parahippocampal gyrus and the temporal lobe (**Figure 2e**).

*Consistency of SNP-level statistics between CLUB-PLS and univariate reference methods*

We used published GWAS data together with the GWIS approach to produce GWAS summary statistics for multivariate brain phenotypes. The target phenotype was the weighted linear combination of the individual brain regions obtained from CLUB-PLS. Across the 416,000 overlapping SNPs, the Spearman rank correlation of -log10(p-values) was higher for CT (0.58-0.83) than for SA (0.37-0.74). Overall, p-values by the two approaches are similar in magnitude as demonstrated by the scatterplots (**Figure 3**). The exception are the SNP associations for SA component 1, where CLUB-PLS p-values are much smaller (**Figure 3A**). To further investigate this discrepancy, we conducted a GWAS on the 33,000 UK Biobank participants with the imaging SA component 1 as the phenotype (as opposed to the GWIS approach). The linear model was adjusted for age (at imaging), $age^2$, ICV, sex and the first 20 genetic principal components for population structure as in our CLUB-PLS analysis. Using GWAS instead of GWIS, there was a very strong correlation between CLUB-PLS p-values and GWAS based p-values (Spearman correlation of $\rho=0.98$; **Figure S1**). Overall, of the 107 different locus-phenotype pairs, 61 showed a genome-wide significant p-value also in the GWIS or the GWAS (only for SA component 2) (**Table 2-3**; **Figure 4**). When lowering the requirement for a validation to genome-wide suggestive ($P<1e-05$), then 85 locus-phenotype pairs could be validated (**Figure S2**).

## Discussion

In this work we introduced CLUB-PLS, which combines federated PLS with the cluster bootstrap technique to enable PLS computations on datasets with arbitrary large sample sizes and feature dimensions along with measures of statistical significance for the projection weights and out-of-sample correlations. As a proof-of-concept CLUB-PLS was applied to an imaging genetics task: finding genetic and imaging associations in 33,000 participants of the UK Biobank. CLUB-PLS discovered 70 loci associated with patterns of surface area and 20 loci associated with patterns of cortical thickness. With the exception for the component modeling asymmetry in cortical thickness, the out-of-sample correlation estimates from the bootstrap approach ranged from 0.075 to 0.18 which are comparable to the predictive capacity of polygenic scores in out-of-sample data reported by Grasby et al. (2020).

The technical validation of CLUB-PLS using GWIS and published summary statistics of regional brain measures demonstrated a good agreement between the approaches. There are various methodological differences that likely contribute to diverging results. Firstly, the used summary statistics adjusted for many covariates (see Smith *et al.* (2021) for details), while in CLUB-PLS we just adjusted for age, $age^2$, sex, ICV and genetic principal components. Secondly, although the datasets were comparable in sample size, there composition was not identical. Thirdly, the GWIS approach only computes an approximation of effect sizes. However, despite these differences, the agreement was substantial as indicated by Spearman correlation values ranging from $\rho$=0.37-0.83. Moreover, conducting a GWAS with SA component 1 as the phenotype, which showed the most notable divergence between CLUB-PLS and GWIS, resulted in nearly identical statistics (Figure S1). In addition, a recent GWAS conducted for principal components of surface area by Shin *et al.* (2020) uncovered a component for general increase or decrease of SA, like our SA component 1. The most significant locus is on chr17 with a p-value ~1e-32, which agrees with our observation (Figure 1, top row) rather than the GWIS result we obtained.

Compared to other multivariate imaging genetics approaches, CLUB-PLS investigates all imaging features and all genetic features at the same time to extract patterns with the strongest co-variation. No other method in the field currently delivers this. The popular MOSTest approach is mass-univariate on the genetics side, and it is unclear what the multivariate imaging patterns looks like that is influenced by the significant genetic loci. These patterns need to be established in post-hoc analyses. Another recent development in imaging genetics, genomic-PC (Fürtjes *et al.*, 2023), operates on summary statistics of GWAS for individual brain regions. Thus, again, the statistics are based on mass-univariate tests. CLUB-PLS is fundamentally different since it is a method for scenarios with access to individual level genetic and imaging data.

The 107 locus-phenotype pairs were mapped to 386 genes (**Table S1**). There were various loci that influenced multiple CLUB-PLS components. For instances, locus 14:59,627,631 (rs2164950) was strongly linked to SA components 2 and 3 as well as CT component 4. This locus has been linked to five genes (*TOMM20L, DACT1, DAAM1, L3HYPDH, RTN1*) and has been previously identified in GWAS using UK Biobank data to be associated with, e.g., regional measures of surface area, cortical thickness and volume (Zhao *et al.*, 2019; Grasby *et al.*, 2020; Smith *et al.*, 2021), cortical folding (van der Meer *et al.*, 2020) and whole brain diffusion (Fan *et al.*, 2022). The locus 17:43,804,619 (Affx-13925359) was linked to component 1 of SA and CT. Interestingly, the SNP tags the common 900-kb inversion polymorphism on 17q21.31 (Stefansson *et al.*, 2005), which covers large number of genes including *MAPT. MAPT* encodes the microtubule-associated protein tau, which is integral to the Alzheimer's disease pathology. Mutations in *MAPT* have been linked to several neurodegenerative disorders including Alzheimer's disease (Myers *et al.*, 2005), frontotemporal dementia (Rademakers *et al.*, 2004), Parkinson's disease (Zabetian *et al.*, 2007) and dementia with lewy bodies (Colom-Cadena *et al.*, 2013). Another locus on chromosome 17 (17:2,574,821; rs12938775) was associated with SA component 1 and CT component 1, which both reflected overall increase in SA and CT, respectively. The locus was linked to *PAFAH1B1,* which is long known to be essential for brain development since loss or mutation of *PAFAH1B1* is linked to lissencephaly (Lo Nigro *et al.*, 1997), a brain malformation characterized by the absence of folds. The list of significant SNP-phenotype pairs contains many instances of loci that have been previously identified with the classic GWAS methodology (Smith *et al.*, 2021) or the MOSTest (van der Meer *et al.*, 2020) using largely the same cohort of UK Biobank participants.

The current proof-of-concept study has several limitations. Firstly, genotyped as opposed to imputed SNPs were used, likewise, the imaging dimensionality was reduced to regions of interest. However, we note that the underlying PLS approach had previously been applied to vertex level imaging data ($p1$= 354,804) and imputed SNPs ($p2$=1,167,126) (Lorenzi *et al.*, 2017, 2018) and therefore CLUB-PLS will be able to handle these dimensions as well. The current limitation to regional imaging features enabled us to conduct a technical validation of the CLUB-PLS results, which would have been computationally impossible using vertex-level phenotypes. Secondly, this analysis represents the discovery part of the typical statistical genetic workflow, and we provided only a technical validation using published univariate GWAS results based on largely the same dataset (Smith *et al.*, 2021). Future work will require a formal validation of the identified associations between loci and multivariate imaging patterns in an intendent cohort, e.g., using the GWIS approach (Nieuwboer *et al.*, 2016) in univariate GWAS for individual brain regions such as done by Grasby *et al.* (2020). Nevertheless, we discovered a series of genetic loci that were previously described in the literature as being associated with cortical morphology. Finally, we would like to emphasize that imaging genetics is only one application field for CLUB-PLS. It is a general approach that can be used in any setting where two high-dimensional modalities are correlated. For instances, CLUB-PLS could be used instead of a sparse CCA approach in lesion to function mapping tasks (Pustina *et al.*, 2018).

## Acknowledgements

AA holds a Medical Research Council eMedLab Medical Bioinformatics Career Development Fellowship. This work was supported by the Medical Research Council (grant number MR/L016311/1). This research has been conducted using the UK Biobank Resource under Application Number 41127.

*Figures*

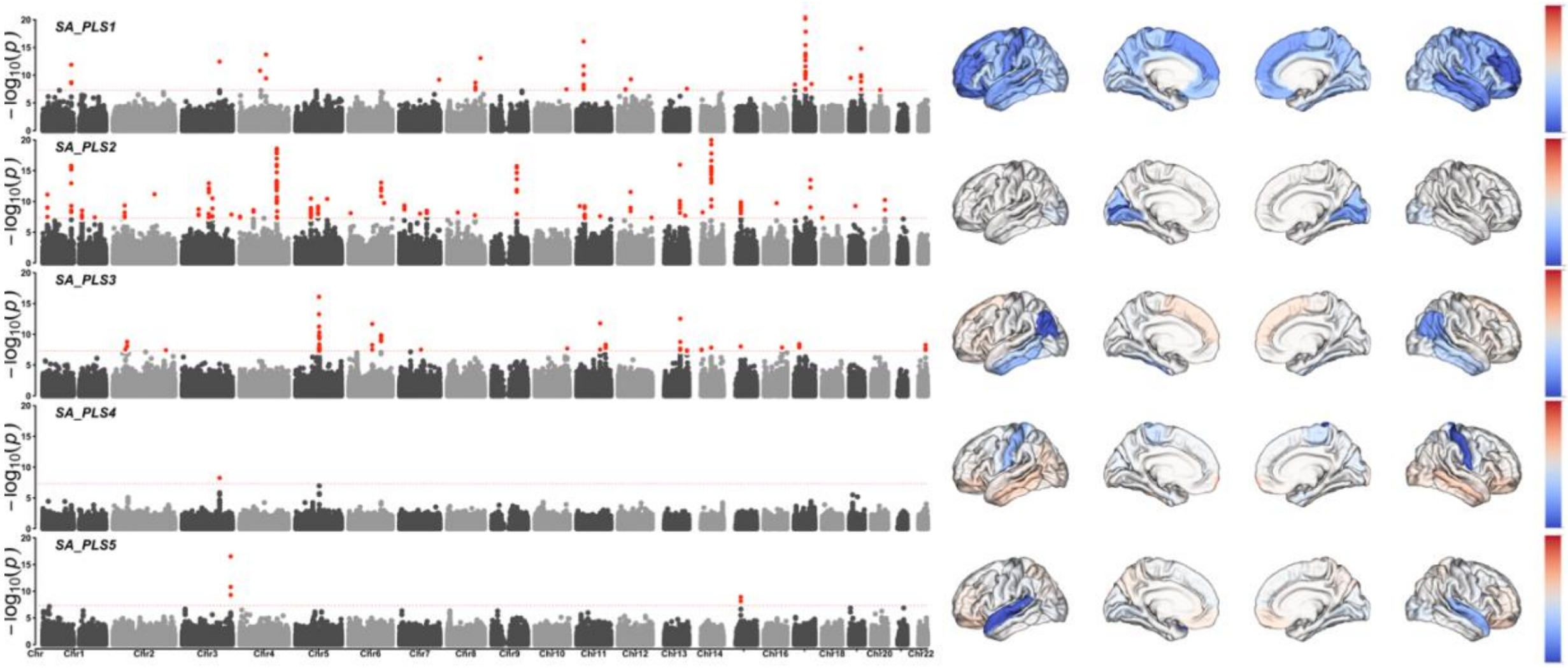

**Figure 1: Matched genetic and imaging associations for Surface Area.** The left part of the plot shows the genetic associations extracted by CLUB-PLS for the imaging phenotype depicted in the right part. Each row corresponds to one CLUB-PLS component. Genetic associations are shown in the form of Manhattan plots with SNPs ordered by chromosomes and positions on the x-axis and the -log10(p-value) on the y-axis. SNPs passing the threshold for genome-wide significance are highlighted in red. The imaging components is colored based on the Wald statistic multiplied by the direction of the effect, with cold and warm colors indicating decreases and increases in SA, respectively.

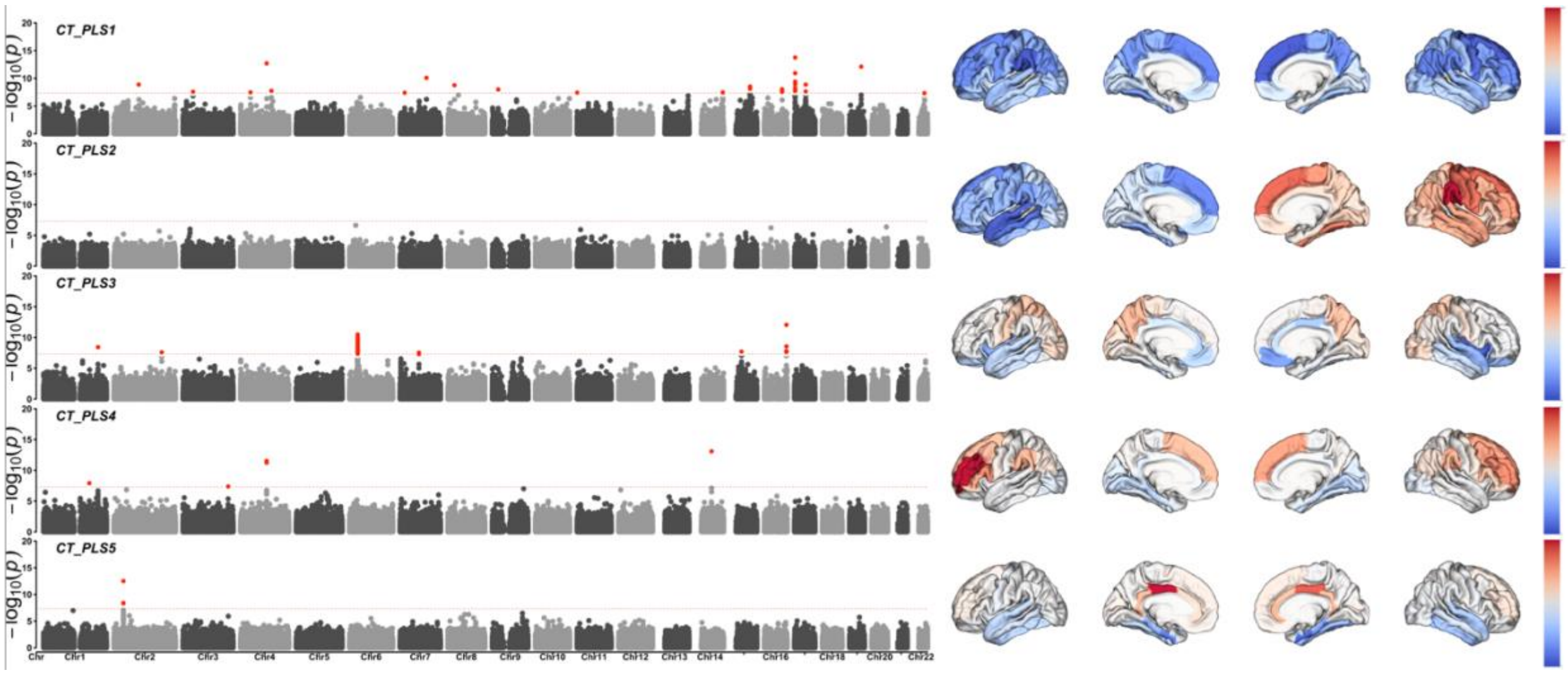

**Figure 2: Matched genetic and imaging associations for Cortical Thickness.** The left part of the plot shows the genetic associations extracted by CLUB-PLS for the imaging phenotype depicted in the right part. Each row corresponds to one CLUB-PLS component. Genetic associations are shown in the form of Manhattan plots with SNPs ordered by chromosomes and positions on the x-axis and the -log10(p-value) on the y-axis. SNPs passing the threshold for genome-wide significance are highlighted in red. The imaging components is colored based on the Wald statistic multiplied by the direction of the effect, with cold and warm colors indicating decreases and increases in CT, respectively.

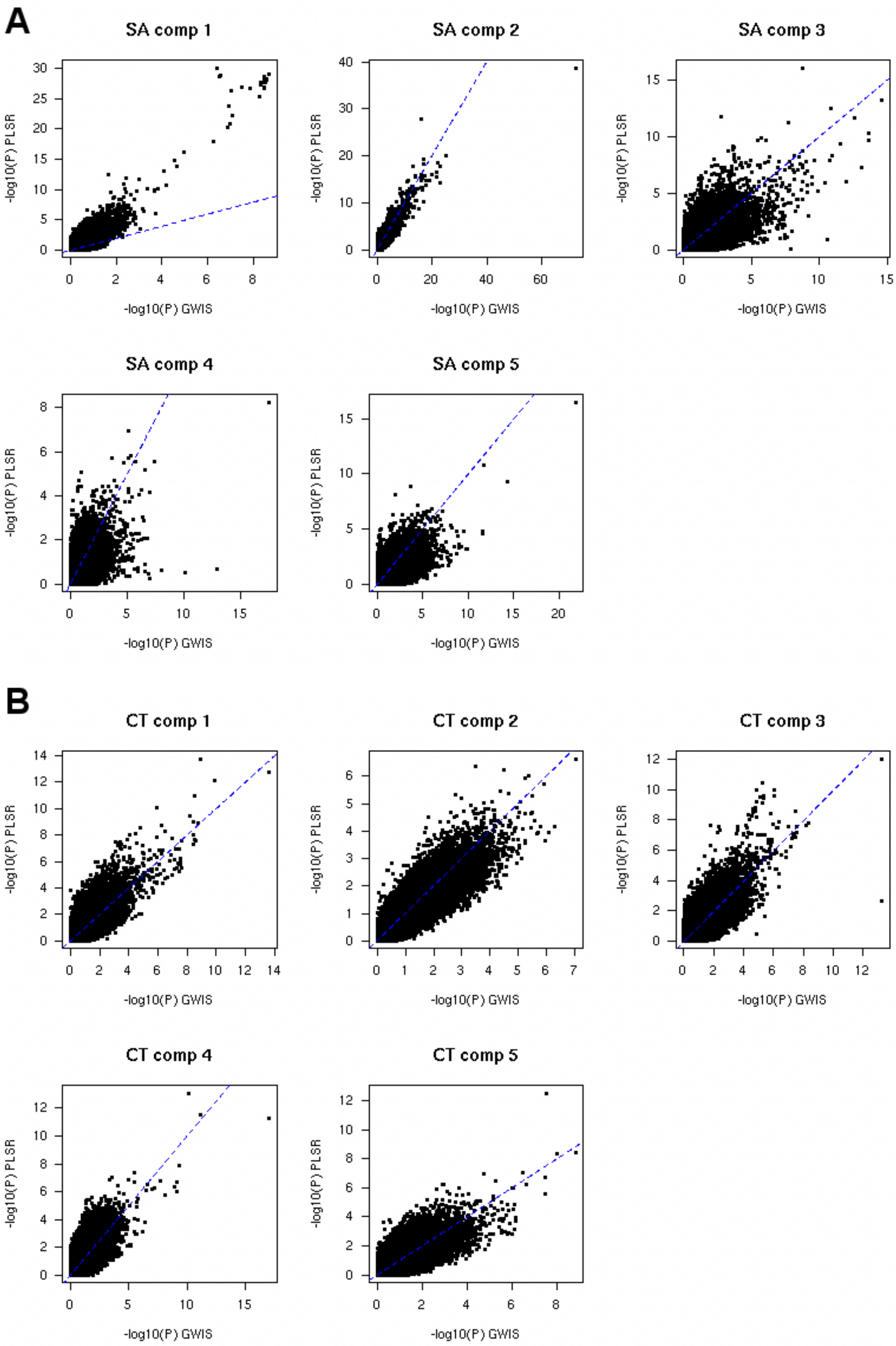

**Figure 3: Correlation between CLUB-PLS and GWIS statistics for SNPs.** The scatter plots depict the -log10(p-values) for GWIS (x-axis) and CLUB-PLS (y-axis) for 416,000 SNPs. The blue dashed line indicates the identity (i.e., the same p-value by both approaches). (**A**) Scatter plots for Surface Area; the Spearman correlation between these two approaches is $\rho$=0.60, 0.74, 0.48, 0.37, and 0.49, for components 1-5, respectively. (**B**) Scatter plots for Cortical Thickness; the Spearman correlation between these two approaches is $\rho$=0.58, 0.83, 0.61, 0.58, and 0.61, for components 1-5, respectively.

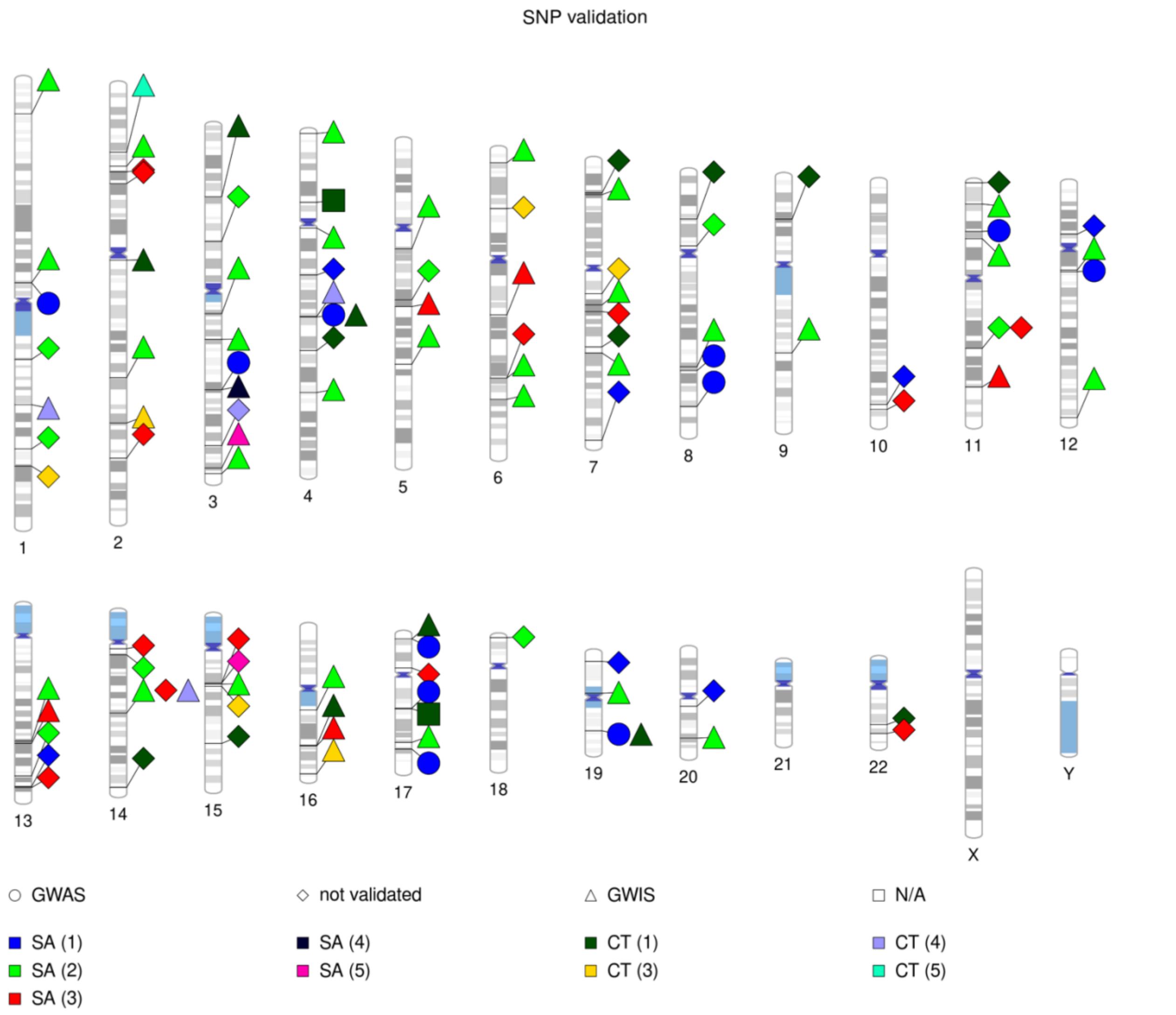

**Figure 4. Technical validation of significant hits identified by CLUB-PLS.** This ideogram depicts the loci of genome-wide significant loci identified with CLUB-PLS. Different colors indicate the different PLS components (SA=Surface Area; CT=Cortical Thickness). The symbols indicate whether the SNP-phenotype combination was validated via GWAS (circle), GWIS (triangle), missing (square) or did not reach the genome-wide significance threshold (diamond). Of the 107 individual SNP-phenotype associations, 61 also exceeded the genome-wide significance threshold (P<5e-8) using the GWAS or GWIS approach and two SNPs were not available in the summary statistics. The remaining 44 SNPs did not pass the genome-wide threshold.

**Tables**

**Table 1: Average out of sample correlations between latent genetics and imaging components from PLS.**

| Component | Cortical Thickness | | Surface Area | |
|---|---|---|---|---|
| | correlation | p-value | correlation | p-value |
| PLS 1 | 0.099 | 1.18e-72 | 0.171 | 5.76e-215 |
| PLS 2 | 0.006 | 0.28 | 0.184 | 3.83e-249 |
| PLS 3 | 0.090 | 2.57e-60 | 0.140 | 4.56e-144 |
| PLS 4 | 0.088 | 9.81e-58 | 0.116 | 3.20e-99 |
| PLS 5 | 0.083 | 1.54e-51 | 0.109 | 9.16e-88 |
| PLS 6 | 0.118 | 1.25e-102 | 0.122 | 1.27e-109 |
| PLS 7 | 0.101 | 1.47e-75 | 0.104 | 5.02e-80 |
| PLS 8 | 0.105 | 1.52e-81 | 0.102 | 4.93e-77 |
| PLS 9 | 0.085 | 5.67e-54 | 0.084 | 9.51e-53 |
| PLS 10 | 0.075 | 2.22e-42 | 0.087 | 1.82e-56 |

Pearson’s correlation coefficient based on the average of 500 bootstrap replicates.

**Table 2. Genome-wide significant loci for the Surface Area components 1-5.**

| Chrom | Position | A1 | A2 | SNP | BETA | SE | P | Validation P |
|---|---|---|---|---|---|---|---|---|
| *Component 1* | | | | | | | | |
| 1 | 113190807 | C | A | rs17030613 | 0.00615 | 0.00086702 | 1.31E-12 | 2.65E-08 |
| 3 | 147054867 | C | T | rs10513309 | 0.00689816 | 0.00094918 | 3.66E-13 | 3.51E-09 |
| 4 | 79832706 | G | C | rs12507099 | -0.0042969 | 0.00063724 | 1.55E-11 | 7.72E-08 |
| 4 | 103188709 | T | C | rs13107325 | -0.0055074 | 0.00071912 | 1.88E-14 | 8.74E-15 |
| 7 | 155778152 | A | G | rs75745313 | 0.00363779 | 0.00058863 | 6.41E-10 | 7.70E-08 |
| 8 | 110603219 | C | G | rs17450520 | -0.005264 | 0.00088009 | 2.21E-09 | 2.50E-08 |
| 8 | 130645692 | G | A | rs55705857 | 0.00482797 | 0.00064695 | 8.48E-14 | 4.96E-14 |
| 10 | 123996970 | A | G | rs61753077 | 0.00271167 | 0.00049217 | 3.60E-08 | 2.83E-06 |
| 11 | 27353309 | G | A | rs10835163 | -0.0071583 | 0.00128152 | 2.33E-08 | 1.01E-09 |
| 12 | 28412415 | TC | T | rs141906323 | -0.0060594 | 0.00109696 | 3.32E-08 | 3.95E-07 |
| 12 | 49379537 | T | G | rs118115924 | -0.0019387 | 0.00031249 | 5.50E-10 | 2.80E-11 |
| 13 | 107095811 | C | T | rs16968624 | -0.0031232 | 0.00056247 | 2.81E-08 | 1.31E-05 |
| 17 | 2574821 | G | A | rs12938775 | 0.00844408 | 0.00144155 | 4.69E-09 | 6.45E-10 |
| 17 | 43804619 | T | C | Affx-13925359 | 0.01381989 | 0.00121225 | 4.17E-30 | 7.72E-33 |
| 17 | 68090207 | T | C | rs11867479 | 0.00793058 | 0.00134683 | 3.90E-09 | 3.40E-09 |
| 19 | 5037351 | A | G | rs263057 | -0.0039072 | 0.00062079 | 3.10E-10 | 2.83E-06 |
| 19 | 46118127 | T | C | rs3816046 | -0.0128144 | 0.001254 | 1.63E-24 | 1.06E-22 |
| 20 | 33565169 | T | C | rs6120778 | 0.00699092 | 0.00127968 | 4.68E-08 | 1.61E-07 |
| *Component 2* | | | | | | | | |
| 1 | 18976489 | C | T | rs9439714 | 0.00899863 | 0.00131447 | 7.60E-12 | 1.52E-12 |
| 1 | 113083439 | T | C | rs10745330 | 0.01212021 | 0.00146881 | 1.56E-16 | 7.42E-18 |
| 1 | 155735012 | C | T | rs2297775 | -0.006725 | 0.00113394 | 3.02E-09 | 1.07E-06 |
| 1 | 205773461 | C | T | rs708729 | 0.00496063 | 0.00090284 | 3.92E-08 | 9.32E-06 |
| 2 | 45170153 | A | G | rs4953152 | -0.0082434 | 0.00132278 | 4.61E-10 | 1.79E-09 |
| 2 | 162891075 | T | C | rs2160927 | 0.00777548 | 0.00113332 | 6.85E-12 | 1.57E-11 |
| 3 | 64546459 | T | C | rs7615657 | 0.00765415 | 0.00127311 | 1.83E-09 | 7.89E-08 |
| 3 | 104735446 | G | A | rs28421781 | 0.00726027 | 0.0009779 | 1.13E-13 | 2.27E-12 |
| 3 | 118959210 | T | C | rs13065867 | 0.00672893 | 0.00101343 | 3.14E-11 | 5.91E-11 |
| 3 | 193810750 | T | C | rs7642977 | -0.0069964 | 0.00123204 | 1.36E-08 | 1.31E-09 |
| 4 | 1003788 | G | C | rs4569707 | -0.008193 | 0.00147727 | 2.92E-08 | 1.12E-08 |
| 4 | 53727952 | G | T | rs76856111 | -0.0058806 | 0.00098692 | 2.55E-09 | 7.97E-10 |
| 4 | 145275039 | A | G | rs6537279 | -0.0100992 | 0.00090919 | 1.15E-28 | 1.18E-16 |
| 5 | 60088250 | C | T | rs6873181 | -0.009681 | 0.00146112 | 3.46E-11 | 4.54E-15 |
| 5 | 88354675 | C | T | rs10037512 | 0.0071891 | 0.00116495 | 6.78E-10 | 7.02E-06 |
| 5 | 124266790 | T | G | rs73302776 | -0.0060895 | 0.00092196 | 3.98E-11 | 1.48E-14 |
| 6 | 7118990 | A | G | rs11755724 | 0.00814659 | 0.00141186 | 7.92E-09 | 6.32E-11 |
| 6 | 127000881 | G | A | rs74580701 | 0.00426351 | 0.00057155 | 8.68E-14 | 2.31E-15 |
| 6 | 138872645 | T | C | rs4481452 | 0.00837295 | 0.00131309 | 1.81E-10 | 6.05E-13 |
| 7 | 18868031 | T | C | rs17349860 | -0.0085778 | 0.00137671 | 4.65E-10 | 4.47E-14 |
| 7 | 80461728 | C | A | rs10251375 | 0.00719593 | 0.0012536 | 9.46E-09 | 6.24E-10 |
| 7 | 107283711 | A | G | rs2701678 | -0.0063404 | 0.00107016 | 3.13E-09 | 5.34E-09 |
| 8 | 41201081 | C | A | rs13254935 | 0.00446019 | 0.0007675 | 6.20E-09 | 1.27E-07 |
| 8 | 108287597 | A | G | rs10090742 | -0.006731 | 0.00119371 | 1.71E-08 | 1.17E-08 |
| 9 | 98231908 | A | G | rs28702657 | 0.00951485 | 0.00115541 | 1.80E-16 | 8.89E-15 |
| 11 | 12072099 | G | A | rs12417334 | 0.00840954 | 0.00135698 | 5.75E-10 | 2.39E-10 |
| 11 | 31547227 | G | T | rs502794 | 0.00802441 | 0.00129591 | 5.94E-10 | 2.10E-11 |
| 11 | 92453759 | T | C | rs1791571 | 0.00693345 | 0.00124413 | 2.50E-08 | 5.96E-08 |
| 12 | 48398080 | T | A | rs3803183 | 0.00707364 | 0.00101302 | 2.90E-12 | 4.16E-12 |
| 12 | 130578130 | G | A | rs7295050 | -0.0069504 | 0.00126885 | 4.31E-08 | 2.87E-08 |
| 13 | 80170160 | G | A | rs2783130 | -0.0134377 | 0.00162031 | 1.10E-16 | 9.20E-21 |
| 13 | 100723564 | C | A | rs2151943 | -0.0034033 | 0.00060652 | 2.01E-08 | 2.56E-06 |
| 14 | 25167511 | A | G | rs12879771 | 0.00621355 | 0.00106756 | 5.87E-09 | 3.80E-05 |
| 14 | 59627631 | A | G | rs2164950 | 0.01583882 | 0.00120491 | 1.81E-39 | 4.77E-73 |
| 15 | 39989140 | A | G | rs8028503 | 0.00591819 | 0.00092046 | 1.28E-10 | 2.48E-10 |
| 16 | 49542037 | A | G | rs12926543 | 0.00825526 | 0.00129573 | 1.88E-10 | 1.74E-12 |
| 17 | 63914750 | G | A | rs1420791 | 0.00476567 | 0.0006273 | 3.03E-14 | 3.37E-13 |
| 18 | 322109 | G | A | rs7236292 | 0.00740222 | 0.00135271 | 4.45E-08 | 3.51E-06 |
| 19 | 24021327 | A | G | rs2194275 | 0.00579419 | 0.00093381 | 5.47E-10 | 1.43E-08 |
| 20 | 52447303 | A | G | rs6022786 | -0.0087427 | 0.00133519 | 5.84E-11 | 1.75E-17 |
| *Component 3* | | | | | | | | |
| 2 | 48292697 | T | C | rs28460586 | -0.0055599 | 0.00100211 | 2.89E-08 | 7.54E-05 |
| 2 | 54858511 | T | C | rs1052788 | 0.00621553 | 0.00103214 | 1.72E-09 | 0.00048741 |
| 2 | 207717431 | C | T | rs6737069 | -0.0058183 | 0.00105754 | 3.76E-08 | 0.0003221 |
| 5 | 92187932 | T | C | rs17669337 | -0.0120169 | 0.00144238 | 8.00E-17 | 1.70E-09 |
| 6 | 92002569 | C | A | rs9345124 | -0.0084063 | 0.00119601 | 2.09E-12 | 2.55E-13 |
| 6 | 126966308 | C | T | rs4549631 | 0.00967394 | 0.00150717 | 1.38E-10 | 1.96E-06 |
| 7 | 84010609 | C | T | rs34822336 | 0.00453894 | 0.00082104 | 3.23E-08 | 5.71E-05 |
| 10 | 126859029 | A | G | rs2681919 | 0.00741684 | 0.00132318 | 2.08E-08 | 2.81E-05 |
| 11 | 92453759 | T | C | rs1791571 | 0.00968131 | 0.00137057 | 1.62E-12 | 0.0016966 |
| 11 | 114186645 | C | T | rs10891647 | -0.0052956 | 0.00090529 | 4.93E-09 | 6.61E-09 |
| 13 | 81428951 | T | C | rs7332768 | 0.00714369 | 0.00097907 | 2.96E-13 | 1.49E-11 |
| 13 | 107642591 | C | T | rs1333186 | -0.0066525 | 0.00121199 | 4.04E-08 | 1.82E-05 |
| 14 | 21578007 | A | G | rs36100359 | 0.0045889 | 0.00082902 | 3.11E-08 | 0.00313837 |
| 14 | 59627631 | A | G | rs2164950 | -0.0062292 | 0.0011002 | 1.50E-08 | 0.0001521 |
| 15 | 39366899 | A | C | rs4924334 | 0.00840011 | 0.00146348 | 9.48E-09 | 0.00641049 |
| 16 | 70661986 | C | T | rs4985412 | -0.0078275 | 0.00138118 | 1.45E-08 | 1.63E-08 |
| 17 | 19812541 | C | T | rs203462 | -0.0070801 | 0.00120314 | 3.99E-09 | 0.00017866 |
| 22 | 43824618 | T | G | rs7290966 | -0.0074383 | 0.00127534 | 5.46E-09 | 4.48E-07 |
| *Component 4* | | | | | | | | |
| 3 | 147175324 | C | T | rs12630408 | 0.00915116 | 0.00156986 | 5.57E-09 | 3.22E-18 |
| *Component 5* | | | | | | | | |
| 3 | 190654424 | C | T | rs13089287 | 0.01345512 | 0.00159297 | 3.00E-17 | 1.77E-22 |
| 15 | 39619456 | A | C | rs8032326 | -0.0070101 | 0.00115448 | 1.26E-09 | 0.00021828 |

Chrom = Chromosome; Position = base pair position on hg19; A1 = affect allele; A2 = reference allele;  BETA = effect size computed by CLUB-PLS; SE = standard error; P = CLUB-PLS p-value; Validation P = GWAS (for component 1) or GWIS p-value

**Table 3: Genome-wide significant loci for the Cortical Thickness components 1-5.**

| Chrom | Position | A1 | A2 | SNP | BETA | SE | P | Validation P |
|---|---|---|---|---|---|---|---|---|
| *Component 1* | | | | | | | | |
| 2 | 97668945 | T | C | rs78783493 | 0.00515147 | 0.00084991 | 1.35E-09 | 1.71E-09 |
| 3 | 39523003 | T | C | rs1768208 | 0.00620018 | 0.00111445 | 2.64E-08 | 2.67E-08 |
| 4 | 39302247 | C | T | rs12500277 | 0.00237978 | 0.00043077 | 3.31E-08 | N/A |
| 4 | 103188709 | T | C | rs13107325 | 0.00587969 | 0.00079968 | 1.94E-13 | 2.52E-14 |
| 4 | 121707714 | T | C | rs12507922 | -0.0081138 | 0.00144149 | 1.82E-08 | 1.11E-07 |
| 7 | 18008582 | G | A | rs34058374 | -0.0070037 | 0.00127528 | 3.98E-08 | 2.78E-05 |
| 7 | 103733264 | A | G | rs10276148 | 0.0083364 | 0.00128321 | 8.22E-11 | 1.35E-06 |
| 8 | 26137132 | A | C | rs73215685 | -0.0030736 | 0.00051054 | 1.74E-09 | 6.39E-07 |
| 9 | 23529626 | C | A | rs156396 | -0.0064789 | 0.00113151 | 1.03E-08 | 9.52E-05 |
| 11 | 305619 | T | C | rs6421984 | 0.00558426 | 0.00101466 | 3.72E-08 | 0.00046413 |
| 14 | 103834316 | A | G | rs2403171 | 0.00553514 | 0.0010022 | 3.33E-08 | 0.00080329 |
| 15 | 75136261 | C | G | rs6938 | -0.0069693 | 0.00117609 | 3.11E-09 | 1.81E-05 |
| 16 | 70148508 | A | G | rs4985397 | 0.00828962 | 0.00144345 | 9.31E-09 | 3.75E-09 |
| 17 | 2546340 | T | C | rs74252325 | -0.0080062 | 0.00104407 | 1.74E-14 | 1.23E-09 |
| 17 | 44109474 | A | G | Affx-13930388 | 0.00814595 | 0.00134359 | 1.34E-09 | N/A |
| 19 | 46118127 | T | C | rs3816046 | 0.00921296 | 0.00128614 | 7.88E-13 | 1.28E-10 |
| 22 | 38491765 | T | G | rs6000996 | -0.0052419 | 0.00095999 | 4.75E-08 | 0.00036129 |
| *Component 3* | | | | | | | | |
| 1 | 215379943 | A | G | rs2027320 | 0.00709248 | 0.00120244 | 3.67E-09 | 1.49E-06 |
| 2 | 188186079 | A | G | rs840584 | -0.007546 | 0.00135485 | 2.55E-08 | 2.63E-08 |
| 6 | 32578970 | C | T | rs502771 | -0.0057444 | 0.00086634 | 3.34E-11 | 5.98E-06 |
| 7 | 74094721 | G | T | rs13227433 | 0.0057065 | 0.00102927 | 2.95E-08 | 0.00433 |
| 15 | 41427864 | T | C | rs561821 | 0.00731508 | 0.00130264 | 1.96E-08 | 1.13E-06 |
| 16 | 87226206 | T | C | rs9933149 | -0.0108802 | 0.00152189 | 8.73E-13 | 6.21E-14 |
| *Component 4* | | | | | | | | |
| 1 | 180962282 | A | G | rs1411478 | -0.0072324 | 0.00126922 | 1.21E-08 | 4.50E-10 |
| 3 | 178072001 | A | G | rs2583481 | 0.00733272 | 0.00133712 | 4.16E-08 | 3.28E-06 |
| 4 | 102926923 | A | G | rs34592089 | 0.00443904 | 0.00063599 | 2.96E-12 | 7.99E-12 |
| 14 | 59627631 | A | G | rs2164950 | 0.00595363 | 0.00079703 | 8.03E-14 | 7.13E-11 |
| *Component 5* | | | | | | | | |
| 2 | 37232014 | A | G | rs2247935 | 0.00808785 | 0.00110935 | 3.09E-13 | 3.10E-08 |

Chrom = Chromosome; Position = base pair position on hg19; A1 = affect allele; A2 = reference allele; BETA = effect size computed by CLUB-PLS; SE = standard error; P = CLUB-PLS p-value; Validation P = GWIS p-value

*Supplementary Material*

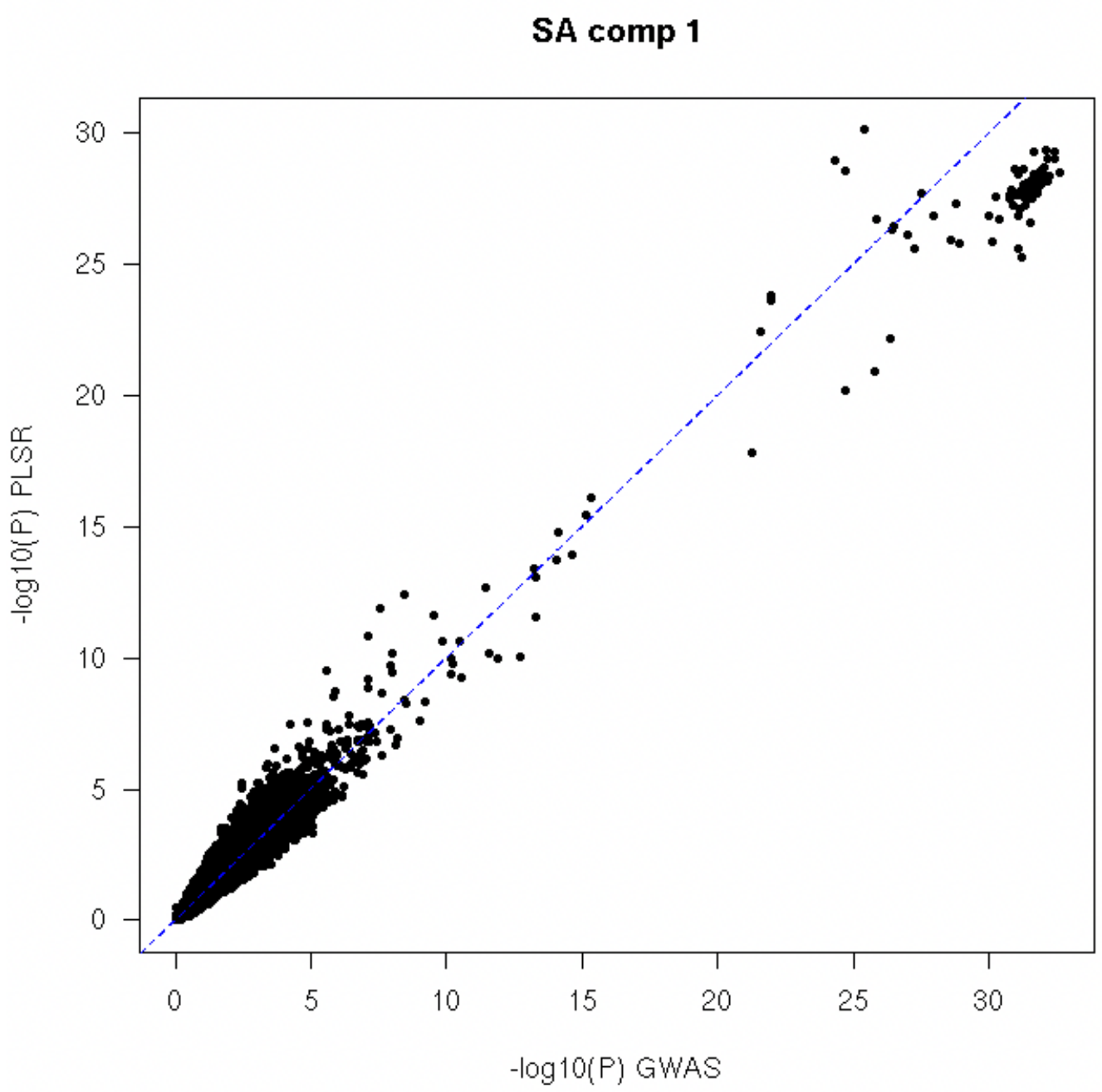

**Figure S1: Correlation between CLUB-PLS and GWAS statistics for SNPs.** The scatter plots depict the -log10(p-values) for GWAS (x-axis) and CLUB-PLS (y-axis) for SA component 1. The blue dashed line indicates the identity (i.e., the same p-value by both approaches). The Spearman correlation was ρ=0.98.

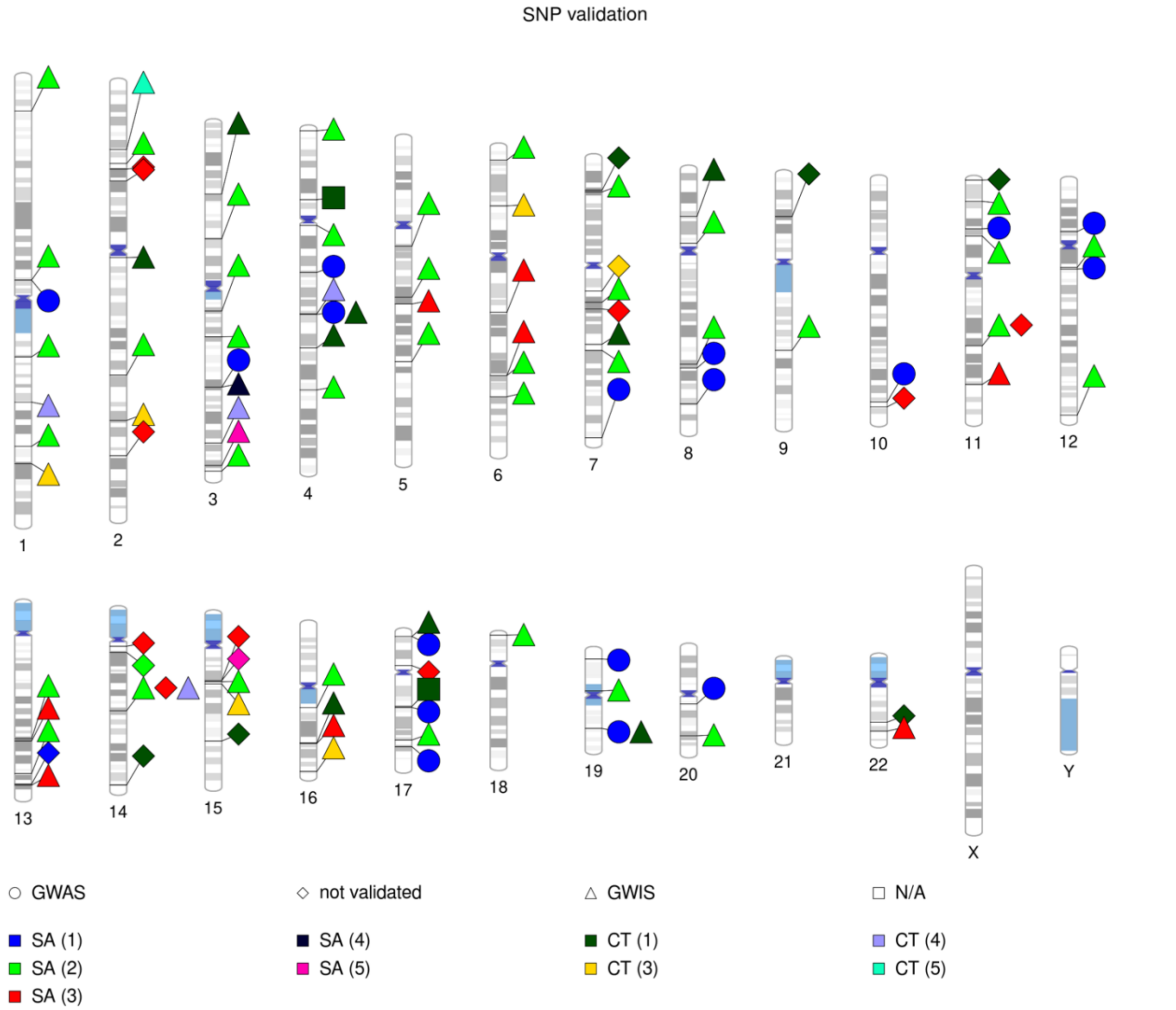

**Figure S2: Technical validation of significant hits identified by CLUB-PLS.** This ideogram depicts the loci of genome-wide significant loci identified with CLUB-PLS. Different colors indicate the different PLS components (SA=Surface Area; CT=Cortical Thickness). The symbols indicate whether the SNP-phenotype combination was validated via GWAS (circle), GWIS (triangle), missing (square) or did not reach the genome-wide suggestive threshold (P<1e-5; diamond). Of the 107 individual SNP-phenotype associations, 85 also exceeded the genome-wide suggestive threshold (P<1e-5) using the GWAS or GWIS approach and two SNPs were not available in the summary statistics. The remaining 20 SNPs did not pass the genome-wide suggestive threshold.

**Table S1: Mapped genes.** Based on the SNP2GENE function in FUMA using positional mapping (within 10kb), eQTL in GTEx Brain Tissues as well as 3D chromatin interaction maps. (File: Table_S1.pdf).

| ensg | symbol | chr | start | end | entrezID | HUGO | eqtlMapminQ | eqtlDirection | ciMap | GenomicLocus |
|---|---|---|---|---|---|---|---|---|---|---|
| ENSG00000009709 | PAX7 | 1 | 18957500 | 19075360 | 5081 | PAX7 | NA | NA | Yes | 1 |
| ENSG00000143110 | C1orf162 | 1 | 112016414 | 112021134 | 128346 | C1orf162 | NA | NA | Yes | 2 |
| ENSG00000143079 | CTTNBP2NL | 1 | 112938803 | 113006078 | 55917 | CTTNBP2NL | NA | NA | Yes | 2 |
| ENSG00000134245 | WNT2B | 1 | 113009163 | 113072787 | 7482 | WNT2B | 1.90E-14 | - | Yes | 2 |
| ENSG00000007341 | ST7L | 1 | 113066140 | 113163447 | 54879 | ST7L | 7.65E-09 | + | Yes | 2 |
| ENSG00000116489 | CAPZA1 | 1 | 113161795 | 113214241 | 829 | CAPZA1 | 9.78E-05 | + | Yes | 2 |
| ENSG00000155363 | MOV10 | 1 | 113215763 | 113243368 | 4343 | MOV10 | NA | NA | Yes | 2 |
| ENSG00000155366 | RHOC | 1 | 113243728 | 113250056 | 389 | RHOC | NA | NA | Yes | 2 |
| ENSG00000271810 | RP11-426L16.10 | 1 | 113245236 | 113254055 | NA | NA | NA | NA | No | 2 |
| ENSG00000184599 | FAM19A3 | 1 | 113263041 | 113269857 | 284467 | FAM19A3 | NA | NA | Yes | 2 |
| ENSG00000155380 | SLC16A1 | 1 | 113454469 | 113499635 | 6566 | SLC16A1 | NA | NA | Yes | 2 |
| ENSG00000081026 | MAGI3 | 1 | 113933371 | 114228545 | 260425 | MAGI3 | NA | NA | Yes | 2 |
| ENSG00000134262 | AP4B1 | 1 | 114437370 | 114447823 | 10717 | AP4B1 | NA | NA | Yes | 2 |
| ENSG00000118655 | DCLRE1B | 1 | 114447763 | 114456708 | 64858 | DCLRE1B | NA | NA | Yes | 2 |
| ENSG00000163349 | HIPK1 | 1 | 114471814 | 114520426 | 204851 | HIPK1 | NA | NA | Yes | 2 |
| ENSG00000197323 | TRIM33 | 1 | 114935399 | 115053781 | 51592 | TRIM33 | NA | NA | Yes | 2 |
| ENSG00000163344 | PMVK | 1 | 154897210 | 154909467 | 10654 | PMVK | 3.03E-09 | NA | No | 3 |
| ENSG00000163348 | PYGO2 | 1 | 154929502 | 154936329 | 90780 | PYGO2 | 0.00635244 | NA | No | 3 |
| ENSG00000160688 | FLAD1 | 1 | 154955814 | 154965587 | 80308 | FLAD1 | NA | NA | Yes | 3 |
| ENSG00000143537 | ADAM15 | 1 | 155023042 | 155035252 | 8751 | ADAM15 | 2.53E-06 | + | No | 3 |
| ENSG00000169242 | EFNA1 | 1 | 155099936 | 155107333 | 1942 | EFNA1 | NA | NA | Yes | 3 |
| ENSG00000169241 | SLC50A1 | 1 | 155107820 | 155111329 | 55974 | SLC50A1 | NA | NA | Yes | 3 |
| ENSG00000179085 | DPM3 | 1 | 155112367 | 155113071 | 54344 | DPM3 | NA | NA | Yes | 3 |
| ENSG00000273088 | RP11-201K10.3 | 1 | 155141885 | 155159748 | NA | NA | NA | NA | Yes | 3 |
| ENSG00000169231 | THBS3 | 1 | 155165379 | 155178842 | 7059 | THBS3 | 2.36E-13 | NA | Yes | 3 |
| ENSG00000173171 | MTX1 | 1 | 155178490 | 155183615 | 4580 | MTX1 | NA | NA | Yes | 3 |
| ENSG00000177628 | GBA | 1 | 155204243 | 155214490 | 2629 | GBA | NA | NA | Yes | 3 |
| ENSG00000160767 | FAM189B | 1 | 155216996 | 155225274 | 10712 | FAM189B | NA | NA | No | 3 |
| ENSG00000116521 | SCAMP3 | 1 | 155225770 | 155232221 | 10067 | SCAMP3 | NA | NA | Yes | 3 |
| ENSG00000176444 | CLK2 | 1 | 155232659 | 155248282 | 1196 | CLK2 | NA | NA | Yes | 3 |
| ENSG00000143630 | HCN3 | 1 | 155247374 | 155259639 | 57657 | HCN3 | 0.0024497 | NA | Yes | 3 |
| ENSG00000143627 | PKLR | 1 | 155259630 | 155271225 | 5313 | PKLR | NA | NA | Yes | 3 |
| ENSG00000160752 | FDPS | 1 | 155278539 | 155290457 | 2224 | FDPS | NA | NA | Yes | 3 |
| ENSG00000160753 | RUSC1 | 1 | 155290687 | 155300905 | 23623 | RUSC1 | NA | NA | Yes | 3 |
| ENSG00000116539 | ASH1L | 1 | 155305059 | 155532598 | 55870 | ASH1L | NA | NA | No | 3 |
| ENSG00000125459 | MSTO1 | 1 | 155579979 | 155718153 | 55154 | MSTO1 | 5.68E-14 | - | No | 3 |
| ENSG00000163374 | YY1AP1 | 1 | 155629237 | 155658791 | 55249 | YY1AP1 | NA | NA | No | 3 |
| ENSG00000132676 | DAP3 | 1 | 155657751 | 155708801 | 7818 | DAP3 | 0.013506 | NA | No | 3 |
| ENSG00000116580 | GON4L | 1 | 155719508 | 155829191 | 54856 | GON4L | NA | NA | Yes | 3 |
| ENSG00000132718 | SYT11 | 1 | 155829300 | 155854990 | 23208 | SYT11 | NA | NA | Yes | 3 |
| ENSG00000143622 | RIT1 | 1 | 155867599 | 155881195 | 6016 | RIT1 | 2.12E-05 | NA | Yes | 3 |
| ENSG00000132680 | KIAA0907 | 1 | 155882834 | 155904191 | 22889 | KIAA0907 | NA | NA | No | 3 |
| ENSG00000173080 | RXFP4 | 1 | 155911480 | 155912625 | 339403 | RXFP4 | NA | NA | Yes | 3 |
| ENSG00000116584 | ARHGEF2 | 1 | 155916630 | 155976861 | 9181 | ARHGEF2 | NA | NA | No | 3 |
| ENSG00000163479 | SSR2 | 1 | 155978839 | 155990750 | 6746 | SSR2 | 0.00235671 | NA | No | 3 |
| ENSG00000160803 | UBQLN4 | 1 | 156005092 | 156023585 | 56893 | UBQLN4 | 0.00264926 | + | Yes | 3 |
| ENSG00000116586 | LAMTOR2 | 1 | 156024543 | 156028301 | 28956 | LAMTOR2 | NA | NA | Yes | 3 |
| ENSG00000132698 | RAB25 | 1 | 156030951 | 156040295 | 57111 | RAB25 | NA | NA | Yes | 3 |
| ENSG00000160785 | SLC25A44 | 1 | 156163880 | 156182587 | 9673 | SLC25A44 | NA | NA | Yes | 3 |
| ENSG00000198715 | C1orf85 | 1 | 156259880 | 156265463 | 112770 | C1orf85 | NA | NA | Yes | 3 |
| ENSG00000189030 | VHLL | 1 | 156268415 | 156269428 | 391104 | VHLL | NA | NA | Yes | 3 |
| ENSG00000163467 | TSACC | 1 | 156307105 | 156316786 | 128229 | TSACC | NA | NA | Yes | 3 |
| ENSG00000135847 | ACBD6 | 1 | 180244515 | 180472089 | 84320 | ACBD6 | NA | NA | Yes | 4 |
| ENSG00000135835 | KIAA1614 | 1 | 180882290 | 180920750 | 57710 | KIAA1614 | 1.81E-06 | - | Yes | 4 |
| ENSG00000234237 | AL162431.1 | 1 | 180941695 | 180949868 | NA | NA | NA | NA | Yes | 4 |
| ENSG00000135823 | STX6 | 1 | 180941861 | 180992047 | 10228 | STX6 | 9.71E-09 | - | No | 4 |
| ENSG00000153029 | MR1 | 1 | 181003067 | 181031074 | 3140 | MR1 | NA | NA | Yes | 4 |
| ENSG00000117280 | RAB7L1 | 1 | 205737114 | 205744588 | 8934 | RAB7L1 | 8.70E-10 | NA | No | 5 |
| ENSG00000133065 | SLC41A1 | 1 | 205758221 | 205782876 | 254428 | SLC41A1 | NA | NA | No | 5 |
| ENSG00000162877 | PM20D1 | 1 | 205797150 | 205819260 | 148811 | PM20D1 | NA | NA | Yes | 5 |
| ENSG00000198049 | AVPR1B | 1 | 206223976 | 206231639 | 553 | AVPR1B | NA | NA | Yes | 5 |
| ENSG00000180667 | YOD1 | 1 | 207217194 | 207226325 | 55432 | YOD1 | NA | NA | Yes | 5 |
| ENSG00000123836 | PFKFB2 | 1 | 207222801 | 207254369 | 5208 | PFKFB2 | NA | NA | Yes | 5 |
| ENSG00000082482 | KCNK2 | 1 | 215179118 | 215410436 | 3776 | KCNK2 | NA | NA | No | 6 |
| ENSG00000268688 | AC007382.1 | 2 | 37068327 | 37068530 | NA | NA | NA | NA | No | 7 |
| ENSG00000115808 | STRN | 2 | 37070783 | 37193615 | 6801 | STRN | NA | NA | No | 7 |
| ENSG00000008869 | HEATR5B | 2 | 37195526 | 37311485 | 54497 | HEATR5B | NA | NA | No | 7 |
| ENSG00000138083 | SIX3 | 2 | 45168902 | 45173216 | 6496 | SIX3 | NA | NA | No | 8 |
| ENSG00000095002 | MSH2 | 2 | 47630108 | 47789450 | 4436 | MSH2 | NA | NA | Yes | 9 |
| ENSG00000116062 | MSH6 | 2 | 47922669 | 48037240 | 2956 | MSH6 | 0.00489396 | NA | No | 9 |
| ENSG00000138081 | FBXO11 | 2 | 48016455 | 48132932 | 80204 | FBXO11 | NA | NA | Yes | 9 |
| ENSG00000170802 | FOXN2 | 2 | 48541776 | 48606433 | 3344 | FOXN2 | 5.46E-52 | NA | No | 9 |
| ENSG00000162869 | PPP1R21 | 2 | 48667737 | 48742525 | 129285 | PPP1R21 | 4.28E-14 | NA | Yes | 9 |
| ENSG00000177994 | C2orf73 | 2 | 54557171 | 54610879 | 129852 | C2orf73 | NA | NA | Yes | 10 |
| ENSG00000115306 | SPTBN1 | 2 | 54683422 | 54896812 | 6711 | SPTBN1 | 2.51E-07 | - | Yes | 10 |
| ENSG00000214595 | EML6 | 2 | 54950636 | 55199157 | 400954 | EML6 | NA | NA | Yes | 10 |
| ENSG00000115310 | RTN4 | 2 | 55199325 | 55339757 | 57142 | RTN4 | NA | NA | Yes | 10 |
| ENSG00000085760 | MTIF2 | 2 | 55463731 | 55496483 | 4528 | MTIF2 | NA | NA | Yes | 10 |
| ENSG00000168754 | FAM178B | 2 | 97541620 | 97684175 | 51252 | FAM178B | NA | NA | No | 11 |
| ENSG00000144199 | FAHD2B | 2 | 97749320 | 97760619 | 151313 | FAHD2B | NA | NA | No | 11 |
| ENSG00000135976 | ANKRD36 | 2 | 97779233 | 97930258 | 375248 | ANKRD36 | NA | NA | No | 11 |
| ENSG00000135940 | COX5B | 2 | 98262503 | 98264846 | 1329 | COX5B | NA | NA | No | 11 |
| ENSG00000115073 | ACTR1B | 2 | 98272431 | 98280570 | 10120 | ACTR1B | 4.16E-26 | + | No | 11 |
| ENSG00000115085 | ZAP70 | 2 | 98330023 | 98356325 | 7535 | ZAP70 | 7.74E-43 | - | No | 11 |
| ENSG00000075568 | TMEM131 | 2 | 98372799 | 98612388 | 23505 | TMEM131 | NA | NA | No | 11 |
| ENSG00000222000 | AC092675.3 | 2 | 98947852 | 98972468 | NA | NA | NA | NA | Yes | 11 |
| ENSG00000144191 | CNGA3 | 2 | 98962618 | 99015064 | 1261 | CNGA3 | NA | NA | Yes | 11 |
| ENSG00000183513 | COA5 | 2 | 99215773 | 99224978 | 493753 | COA5 | NA | NA | Yes | 11 |
| ENSG00000115446 | UNC50 | 2 | 99225042 | 99234978 | 25972 | UNC50 | NA | NA | Yes | 11 |
| ENSG00000071073 | MGAT4A | 2 | 99235569 | 99347589 | 11320 | MGAT4A | NA | NA | Yes | 11 |
| ENSG00000158411 | MITD1 | 2 | 99777890 | 99797521 | 129531 | MITD1 | NA | NA | Yes | 11 |
| ENSG00000185414 | MRPL30 | 2 | 99797542 | 99814089 | 51263 | MRPL30 | NA | NA | Yes | 11 |
| ENSG00000115233 | PSMD14 | 2 | 162164549 | 162268228 | 10213 | PSMD14 | NA | NA | Yes | 12 |
| ENSG00000136535 | TBR1 | 2 | 162272605 | 162282381 | 10716 | TBR1 | NA | NA | Yes | 12 |
| ENSG00000144290 | SLC4A10 | 2 | 162280843 | 162841792 | 57282 | SLC4A10 | NA | NA | No | 12 |
| ENSG00000197635 | DPP4 | 2 | 162848751 | 162931052 | 1803 | DPP4 | NA | NA | No | 12 |
| ENSG00000213953 | AC018867.1 | 2 | 187361840 | 187365393 | NA | NA | NA | NA | Yes | 13 |
| ENSG00000268846 | AC018867.2 | 2 | 187367219 | 187367356 | NA | NA | NA | NA | Yes | 13 |
| ENSG00000138448 | ITGAV | 2 | 187454792 | 187545628 | 3685 | ITGAV | NA | NA | Yes | 13 |
| ENSG00000144369 | FAM171B | 2 | 187558698 | 187630685 | 165215 | FAM171B | NA | NA | Yes | 13 |
| ENSG00000163012 | ZSWIM2 | 2 | 187692562 | 187713935 | 151112 | ZSWIM2 | 0.0346681 | NA | Yes | 13 |
| ENSG00000064989 | CALCRL | 2 | 188207856 | 188313187 | 10203 | CALCRL | 2.84E-05 | + | Yes | 13 |
| ENSG00000003436 | TFPI | 2 | 188328957 | 188430487 | 7035 | TFPI | NA | NA | No | 13 |
| ENSG00000023228 | NDUFS1 | 2 | 206979541 | 207024327 | 4719 | NDUFS1 | NA | NA | Yes | 14 |
| ENSG00000114942 | EEF1B2 | 2 | 207024309 | 207027652 | 619569 | EEF1B2 | NA | NA | Yes | 14 |
| ENSG00000204186 | ZDBF2 | 2 | 207139387 | 207179148 | 57683 | ZDBF2 | NA | NA | Yes | 14 |
| ENSG00000138400 | MDH1B | 2 | 207602487 | 207630271 | 130752 | MDH1B | 1.96E-21 | NA | Yes | 14 |
| ENSG00000118246 | FASTKD2 | 2 | 207630081 | 207657233 | 22868 | FASTKD2 | NA | NA | Yes | 14 |
| ENSG00000118263 | KLF7 | 2 | 207938861 | 208031991 | 8609 | KLF7 | NA | NA | Yes | 14 |
| ENSG00000118260 | CREB1 | 2 | 208394461 | 208468155 | 1385 | CREB1 | NA | NA | Yes | 14 |
| ENSG00000163249 | CCNYL1 | 2 | 208576264 | 208626563 | 151195 | CCNYL1 | NA | NA | Yes | 14 |
| ENSG00000168028 | RPSA | 3 | 39448180 | 39454033 | 574040 | RPSA | 1.12E-13 | + | No | 15 |
| ENSG00000168314 | MOBP | 3 | 39508689 | 39570970 | 4336 | MOBP | NA | NA | No | 15 |
| ENSG00000114784 | EIF1B | 3 | 40351175 | 40353915 | 10289 | EIF1B | NA | NA | Yes | 15 |
| ENSG00000163638 | ADAMTS9 | 3 | 64501333 | 64673676 | 56999 | ADAMTS9 | NA | NA | Yes | 16 |
| ENSG00000170017 | ALCAM | 3 | 105085753 | 105295744 | 214 | ALCAM | NA | NA | Yes | 17 |
| ENSG00000251012 | RP11-484M3.5 | 3 | 118866222 | 118906654 | NA | NA | NA | NA | No | 18 |
| ENSG00000114638 | UPK1B | 3 | 118892364 | 118924000 | 7348 | UPK1B | 0.00764953 | - | No | 18 |
| ENSG00000121578 | B4GALT4 | 3 | 118930579 | 118959950 | 8702 | B4GALT4 | NA | NA | Yes | 18 |
| ENSG00000176142 | TMEM39A | 3 | 119148347 | 119187677 | 55254 | TMEM39A | NA | NA | Yes | 18 |
| ENSG00000163389 | POGLUT1 | 3 | 119187785 | 119213555 | 56983 | POGLUT1 | NA | NA | Yes | 18 |
| ENSG00000113845 | TIMMDC1 | 3 | 119217379 | 119243937 | 51300 | TIMMDC1 | NA | NA | Yes | 18 |
| ENSG00000144843 | ADPRH | 3 | 119298115 | 119308792 | 141 | ADPRH | NA | NA | Yes | 18 |
| ENSG00000174963 | ZIC4 | 3 | 147103833 | 147124647 | 84107 | ZIC4 | NA | NA | Yes | 19 |
| ENSG00000152977 | ZIC1 | 3 | 147111209 | 147228080 | 7545 | ZIC1 | NA | NA | Yes | 19 |
| ENSG00000144891 | AGTR1 | 3 | 148415571 | 148460795 | 185 | AGTR1 | NA | NA | Yes | 19 |
| ENSG00000153002 | CPB1 | 3 | 148508889 | 148577974 | 1360 | CPB1 | NA | NA | Yes | 19 |
| ENSG00000177565 | TBL1XR1 | 3 | 176737143 | 176915261 | 79718 | TBL1XR1 | NA | NA | Yes | 20 |
| ENSG00000197584 | KCNMB2 | 3 | 177990720 | 178562217 | 10242 | KCNMB2 | NA | NA | No | 20 |
| ENSG00000205835 | GMNC | 3 | 190570666 | 190610218 | 647309 | GMNC | NA | NA | No | 21 |
| ENSG00000114315 | HES1 | 3 | 193853934 | 193856521 | 3280 | HES1 | NA | NA | Yes | 22 |
| ENSG00000145217 | SLC26A1 | 4 | 972861 | 987228 | 10861 | SLC26A1 | 3.90E-15 | + | No | 23 |
| ENSG00000127415 | IDUA | 4 | 980785 | 998316 | 3425 | IDUA | NA | NA | No | 23 |
| ENSG00000127418 | FGFRL1 | 4 | 1003724 | 1020685 | 53834 | FGFRL1 | NA | NA | No | 23 |
| ENSG00000109790 | KLHL5 | 4 | 39046659 | 39128477 | 51088 | KLHL5 | NA | NA | No | 24 |
| ENSG00000035928 | RFC1 | 4 | 39289076 | 39367995 | 5981 | RFC1 | NA | NA | Yes | 24 |
| ENSG00000109189 | USP46 | 4 | 53457138 | 53525502 | 64854 | USP46 | 0.00605346 | NA | Yes | 25 |
| ENSG00000226887 | ERVMER34-1 | 4 | 53588785 | 53617807 | 100288413 | ERVMER34-1 | NA | NA | No | 25 |
| ENSG00000128045 | RASL11B | 4 | 53728457 | 53733000 | 65997 | RASL11B | NA | NA | No | 25 |
| ENSG00000184178 | SCFD2 | 4 | 53739149 | 54232242 | 152579 | SCFD2 | NA | NA | No | 25 |
| ENSG00000145358 | DDIT4L | 4 | 101107027 | 101111939 | 115265 | DDIT4L | NA | NA | Yes | 26 |
| ENSG00000153064 | BANK1 | 4 | 102332443 | 102995969 | 55024 | BANK1 | NA | NA | No | 26 |
| ENSG00000138821 | SLC39A8 | 4 | 103172198 | 103352415 | 64116 | SLC39A8 | NA | NA | No | 26 |
| ENSG00000138738 | PRDM5 | 4 | 121606074 | 121844025 | 11107 | PRDM5 | 0.00107124 | - | No | 27 |
| ENSG00000173376 | NDNF | 4 | 121956768 | 121994176 | 79625 | NDNF | NA | NA | Yes | 27 |
| ENSG00000035499 | DEPDC1B | 5 | 59892739 | 59996017 | 55789 | DEPDC1B | 9.55E-07 | + | No | 29 |
| ENSG00000164181 | ELOVL7 | 5 | 60047618 | 60140216 | 79993 | ELOVL7 | NA | NA | No | 29 |
| ENSG00000049167 | ERCC8 | 5 | 60169658 | 60240900 | 1161 | ERCC8 | 0.0348806 | NA | No | 29 |
| ENSG00000153140 | CETN3 | 5 | 89688078 | 89705603 | 1070 | CETN3 | NA | NA | Yes | 30 |
| ENSG00000175745 | NR2F1 | 5 | 92919043 | 92930321 | 7025 | NR2F1 | NA | NA | Yes | 31 |
| ENSG00000185261 | KIAA0825 | 5 | 93488671 | 93954309 | 285600 | KIAA0825 | NA | NA | Yes | 31 |
| ENSG00000133302 | ANKRD32 | 5 | 93954052 | 94075141 | 84250 | ANKRD32 | NA | NA | Yes | 31 |
| ENSG00000168916 | ZNF608 | 5 | 123972608 | 124084500 | 57507 | ZNF608 | NA | NA | Yes | 32 |
| ENSG00000155324 | GRAMD3 | 5 | 125695824 | 125832186 | 65983 | GRAMD3 | NA | NA | Yes | 32 |
| ENSG00000124782 | RREB1 | 6 | 7107830 | 7252213 | 6239 | RREB1 | 0.0210277 | - | No | 33 |
| ENSG00000203760 | CENPW | 6 | 126661320 | 126670021 | 387103 | CENPW | 0.045306 | NA | No | 35 |
| ENSG00000146374 | RSPO3 | 6 | 127439749 | 127518910 | 84870 | RSPO3 | NA | NA | Yes | 35 |
| ENSG00000016402 | IL20RA | 6 | 137321108 | 137366298 | 53832 | IL20RA | NA | NA | Yes | 36 |
| ENSG00000254440 | PBOV1 | 6 | 138537129 | 138539627 | 59351 | PBOV1 | NA | NA | Yes | 36 |
| ENSG00000262543 | RP3-422G23.4 | 6 | 138699042 | 138704212 | NA | NA | NA | NA | Yes | 36 |
| ENSG00000051620 | HEBP2 | 6 | 138724668 | 138743334 | 23593 | HEBP2 | NA | NA | Yes | 36 |
| ENSG00000135540 | NHSL1 | 6 | 138743180 | 139013708 | 57224 | NHSL1 | NA | NA | No | 36 |
| ENSG00000203734 | ECT2L | 6 | 139117063 | 139225207 | 345930 | ECT2L | NA | NA | Yes | 36 |
| ENSG00000146386 | ABRACL | 6 | 139349819 | 139364439 | 58527 | ABRACL | NA | NA | Yes | 36 |
| ENSG00000112406 | HECA | 6 | 139456249 | 139501939 | 51696 | HECA | NA | NA | Yes | 36 |
| ENSG00000164440 | TXLNB | 6 | 139561198 | 139613276 | 167838 | TXLNB | NA | NA | Yes | 36 |
| ENSG00000164442 | CITED2 | 6 | 139693393 | 139695757 | 10370 | CITED2 | NA | NA | Yes | 36 |
| ENSG00000071189 | SNX13 | 7 | 17830385 | 17980124 | 23161 | SNX13 | NA | NA | No | 37 |
| ENSG00000048052 | HDAC9 | 7 | 18126572 | 19042039 | 9734 | HDAC9 | NA | NA | No | 38 |
| ENSG00000122691 | TWIST1 | 7 | 19060614 | 19157295 | 7291 | TWIST1 | NA | NA | Yes | 38 |
| ENSG00000106683 | LIMK1 | 7 | 73497263 | 73536855 | 3984 | LIMK1 | NA | NA | Yes | 39 |
| ENSG00000106682 | EIF4H | 7 | 73588575 | 73611431 | 7458 | EIF4H | NA | NA | Yes | 39 |
| ENSG00000049541 | RFC2 | 7 | 73645829 | 73668774 | 5982 | RFC2 | NA | NA | Yes | 39 |
| ENSG00000106665 | CLIP2 | 7 | 73703805 | 73820273 | 7461 | CLIP2 | NA | NA | Yes | 39 |
| ENSG00000006704 | GTF2IRD1 | 7 | 73868120 | 74016931 | 9569 | GTF2IRD1 | NA | NA | Yes | 39 |
| ENSG00000077809 | GTF2I | 7 | 74071994 | 74175026 | 2969 | GTF2I | NA | NA | No | 39 |
| ENSG00000158517 | NCF1 | 7 | 74188309 | 74203659 | 653361 | NCF1 | NA | NA | No | 39 |
| ENSG00000196275 | GTF2IRD2 | 7 | 74210483 | 74267847 | 84163 | GTF2IRD2 | 6.51E-38 | + | No | 39 |
| ENSG00000174374 | WBSCR16 | 7 | 74441226 | 74490064 | 653375 | WBSCR16 | NA | NA | Yes | 39 |
| ENSG00000075223 | SEMA3C | 7 | 80371854 | 80551675 | 10512 | SEMA3C | NA | NA | Yes | 40 |
| ENSG00000075213 | SEMA3A | 7 | 83585093 | 84122040 | 10371 | SEMA3A | NA | NA | No | 41 |
| ENSG00000164815 | ORC5 | 7 | 103766788 | 103848495 | 5001 | ORC5 | NA | NA | No | 42 |
| ENSG00000105851 | PIK3CG | 7 | 106505723 | 106547590 | 5294 | PIK3CG | NA | NA | Yes | 43 |
| ENSG00000005249 | PRKAR2B | 7 | 106685094 | 106802256 | 5577 | PRKAR2B | NA | NA | Yes | 43 |
| ENSG00000105856 | HBP1 | 7 | 106809406 | 106842974 | 26959 | HBP1 | NA | NA | Yes | 43 |
| ENSG00000164597 | COG5 | 7 | 106842000 | 107204959 | 10466 | COG5 | 0.00739835 | NA | No | 43 |
| ENSG00000172209 | GPR22 | 7 | 107110463 | 107116098 | 2845 | GPR22 | NA | NA | No | 43 |
| ENSG00000105865 | DUS4L | 7 | 107203929 | 107218906 | 11062 | DUS4L | NA | NA | No | 43 |
| ENSG00000075790 | BCAP29 | 7 | 107220422 | 107269615 | 55973 | BCAP29 | 0.00838518 | + | Yes | 43 |
| ENSG00000091137 | SLC26A4 | 7 | 107301080 | 107358254 | 5172 | SLC26A4 | NA | NA | No | 43 |
| ENSG00000105879 | CBLL1 | 7 | 107384142 | 107402112 | 79872 | CBLL1 | NA | NA | Yes | 43 |
| ENSG00000091140 | DLD | 7 | 107531415 | 107572175 | 1738 | DLD | NA | NA | Yes | 43 |
| ENSG00000104756 | KCTD9 | 8 | 25285366 | 25315992 | 54793 | KCTD9 | NA | NA | Yes | 44 |
| ENSG00000184661 | CDCA2 | 8 | 25316513 | 25365436 | 157313 | CDCA2 | NA | NA | Yes | 44 |
| ENSG00000221818 | EBF2 | 8 | 25699246 | 25902913 | 64641 | EBF2 | NA | NA | Yes | 44 |
| ENSG00000221914 | PPP2R2A | 8 | 26149007 | 26230196 | 5520 | PPP2R2A | NA | NA | Yes | 44 |
| ENSG00000104765 | BNIP3L | 8 | 26240414 | 26363152 | 665 | BNIP3L | NA | NA | Yes | 44 |
| ENSG00000104332 | SFRP1 | 8 | 41119481 | 41167016 | 6422 | SFRP1 | 0.00050797 | + | No | 45 |
| ENSG00000154188 | ANGPT1 | 8 | 108261721 | 108510283 | 284 | ANGPT1 | NA | NA | Yes | 46 |
| ENSG00000107105 | ELAVL2 | 9 | 23690102 | 23826335 | 1993 | ELAVL2 | NA | NA | Yes | 47 |
| ENSG00000158169 | FANCC | 9 | 97861336 | 98079991 | 2176 | FANCC | 7.57E-07 | - | Yes | 48 |
| ENSG00000185920 | PTCH1 | 9 | 98205262 | 98279339 | 5727 | PTCH1 | NA | NA | Yes | 48 |
| ENSG00000268926 | DKFZP434H0512 | 9 | 98534605 | 98537013 | NA | NA | NA | NA | Yes | 48 |
| ENSG00000182150 | ERCC6L2 | 9 | 98637983 | 98776842 | 101928170 | ERCC6L2 | NA | NA | Yes | 48 |
| ENSG00000130956 | HABP4 | 9 | 99212483 | 99253618 | 22927 | HABP4 | NA | NA | Yes | 48 |
| ENSG00000175029 | CTBP2 | 10 | 126676421 | 126849739 | 1488 | CTBP2 | NA | NA | No | 49 |
| ENSG00000142102 | ATHL1 | 11 | 289135 | 296107 | 80162 | ATHL1 | 1.63E-19 | + | No | 50 |
| ENSG00000206013 | IFITM5 | 11 | 298200 | 299526 | 387733 | IFITM5 | NA | NA | No | 50 |
| ENSG00000185201 | IFITM2 | 11 | 307631 | 315272 | 10581 | IFITM2 | 3.06E-08 | - | No | 50 |
| ENSG00000185885 | IFITM1 | 11 | 313506 | 315272 | 8519 | IFITM1 | NA | NA | No | 50 |
| ENSG00000142089 | IFITM3 | 11 | 319669 | 327537 | 10410 | IFITM3 | 1.85E-09 | - | No | 50 |
| ENSG00000133816 | MICAL2 | 11 | 12115543 | 12285334 | 9645 | MICAL2 | 3.45E-21 | - | No | 51 |
| ENSG00000133808 | MICALCL | 11 | 12297627 | 12380691 | 84953 | MICALCL | NA | NA | Yes | 51 |
| ENSG00000170959 | DCDC1 | 11 | 30851916 | 31391357 | 100506627 | DCDC1 | NA | NA | Yes | 52 |
| ENSG00000170946 | DNAJC24 | 11 | 31391387 | 31453396 | 120526 | DNAJC24 | NA | NA | Yes | 52 |
| ENSG00000148950 | IMMP1L | 11 | 31453948 | 31531192 | 196294 | IMMP1L | NA | NA | Yes | 52 |
| ENSG00000109911 | ELP4 | 11 | 31531297 | 31805546 | 26610 | ELP4 | NA | NA | Yes | 52 |
| ENSG00000007372 | PAX6 | 11 | 31806340 | 31839509 | 5080 | PAX6 | NA | NA | Yes | 52 |
| ENSG00000049449 | RCN1 | 11 | 31833939 | 32127301 | 440034 | RCN1 | NA | NA | Yes | 52 |
| ENSG00000184937 | WT1 | 11 | 32409321 | 32457176 | 7490 | WT1 | NA | NA | Yes | 52 |
| ENSG00000165323 | FAT3 | 11 | 92085262 | 92629618 | 120114 | FAT3 | NA | NA | No | 53 |
| ENSG00000166741 | NNMT | 11 | 114128509 | 114184007 | 101928916 | NNMT | NA | NA | No | 54 |
| ENSG00000111371 | SLC38A1 | 12 | 46576846 | 46663800 | 81539 | SLC38A1 | NA | NA | Yes | 55 |
| ENSG00000005175 | RPAP3 | 12 | 48057070 | 48099844 | 79657 | RPAP3 | NA | NA | Yes | 55 |
| ENSG00000268069 | AC004466.1 | 12 | 48178706 | 48179787 | NA | NA | NA | NA | Yes | 55 |
| ENSG00000134291 | TMEM106C | 12 | 48357352 | 48362661 | 79022 | TMEM106C | NA | NA | Yes | 55 |
| ENSG00000139219 | COL2A1 | 12 | 48366748 | 48398269 | 1280 | COL2A1 | NA | NA | No | 55 |
| ENSG00000257955 | RP1-228P16.5 | 12 | 48413554 | 48419165 | NA | NA | NA | NA | Yes | 55 |
| ENSG00000079387 | SENP1 | 12 | 48436681 | 48500091 | 29843 | SENP1 | NA | NA | Yes | 55 |
| ENSG00000152556 | PFKM | 12 | 48498922 | 48540187 | 5213 | PFKM | 1.08E-11 | + | Yes | 55 |
| ENSG00000177981 | ASB8 | 12 | 48541571 | 48574996 | 140461 | ASB8 | 4.72E-11 | NA | Yes | 55 |
| ENSG00000177875 | C12orf68 | 12 | 48577366 | 48579709 | 387856 | C12orf68 | NA | NA | Yes | 55 |
| ENSG00000269514 | DKFZP779L1853 | 12 | 48592170 | 48595814 | NA | NA | NA | NA | Yes | 55 |
| ENSG00000172640 | OR10AD1 | 12 | 48596081 | 48597170 | 121275 | OR10AD1 | NA | NA | Yes | 55 |
| ENSG00000177627 | C12orf54 | 12 | 48876286 | 48890295 | 121273 | C12orf54 | 0.00111042 | NA | No | 55 |
| ENSG00000060709 | RIMBP2 | 12 | 130880682 | 131200826 | 23504 | RIMBP2 | NA | NA | Yes | 56 |
| ENSG00000139746 | RBM26 | 13 | 79885962 | 79980612 | 64062 | RBM26 | 5.69E-06 | + | No | 57 |
| ENSG00000175198 | PCCA | 13 | 100741269 | 101182686 | 5095 | PCCA | NA | NA | No | 59 |
| ENSG00000125266 | EFNB2 | 13 | 107142079 | 107187462 | 1948 | EFNB2 | NA | NA | Yes | 60 |
| ENSG00000134884 | ARGLU1 | 13 | 107194021 | 107220512 | 55082 | ARGLU1 | NA | NA | Yes | 60 |
| ENSG00000100814 | CCNB1IP1 | 14 | 20779527 | 20801471 | 57820 | CCNB1IP1 | NA | NA | Yes | 61 |
| ENSG00000165782 | TMEM55B | 14 | 20925878 | 20929771 | 90809 | TMEM55B | NA | NA | Yes | 61 |
| ENSG00000198805 | PNP | 14 | 20937113 | 20945253 | 4860 | PNP | NA | NA | Yes | 61 |
| ENSG00000165794 | SLC39A2 | 14 | 21467414 | 21470030 | 29986 | SLC39A2 | NA | NA | Yes | 61 |
| ENSG00000165795 | NDRG2 | 14 | 21484922 | 21539031 | 57447 | NDRG2 | NA | NA | No | 61 |
| ENSG00000255472 | RP11-998D10.1 | 14 | 21500977 | 21502054 | NA | NA | NA | NA | Yes | 61 |
| ENSG00000206150 | RNASE13 | 14 | 21500979 | 21502944 | 440163 | RNASE13 | NA | NA | Yes | 61 |
| ENSG00000165799 | RNASE7 | 14 | 21510385 | 21512393 | 84659 | RNASE7 | NA | NA | Yes | 61 |
| ENSG00000173431 | RNASE8 | 14 | 21525981 | 21526614 | 122665 | RNASE8 | NA | NA | Yes | 61 |
| ENSG00000165801 | ARHGEF40 | 14 | 21538429 | 21558399 | 55701 | ARHGEF40 | NA | NA | No | 61 |
| ENSG00000165804 | ZNF219 | 14 | 21558205 | 21572881 | 51222 | ZNF219 | NA | NA | Yes | 61 |
| ENSG00000232070 | TMEM253 | 14 | 21567096 | 21571883 | 643382 | TMEM253 | NA | NA | Yes | 61 |
| ENSG00000129562 | DAD1 | 14 | 23033805 | 23058175 | 1603 | DAD1 | NA | NA | Yes | 61 |
| ENSG00000196860 | TOMM20L | 14 | 58862634 | 58875419 | 387990 | TOMM20L | NA | NA | Yes | 63 |
| ENSG00000165617 | DACT1 | 14 | 59100685 | 59115039 | 51339 | DACT1 | NA | NA | Yes | 63 |
| ENSG00000100592 | DAAM1 | 14 | 59655364 | 59838123 | 23002 | DAAM1 | NA | NA | Yes | 63 |
| ENSG00000126790 | L3HYPDH | 14 | 59927081 | 59951148 | 112849 | L3HYPDH | 1.39E-21 | - | No | 63 |
| ENSG00000139970 | RTN1 | 14 | 60062694 | 60337684 | 6252 | RTN1 | NA | NA | Yes | 63 |
| ENSG00000100664 | EIF5 | 14 | 103799881 | 103811362 | 1983 | EIF5 | NA | NA | Yes | 64 |
| ENSG00000075413 | MARK3 | 14 | 103851729 | 103970168 | 4140 | MARK3 | NA | NA | Yes | 64 |
| ENSG00000166165 | CKB | 14 | 103985996 | 103989448 | 1152 | CKB | NA | NA | Yes | 64 |
| ENSG00000166166 | TRMT61A | 14 | 103995521 | 104003410 | 115708 | TRMT61A | 8.83E-05 | NA | No | 64 |
| ENSG00000175779 | C15orf53 | 15 | 38988799 | 38992239 | 400359 | C15orf53 | NA | NA | Yes | 65 |
| ENSG00000137801 | THBS1 | 15 | 39873280 | 39891667 | 7057 | THBS1 | NA | NA | Yes | 65 |
| ENSG00000150667 | FSIP1 | 15 | 39892232 | 40075031 | 161835 | FSIP1 | NA | NA | No | 65 |
| ENSG00000128944 | KNSTRN | 15 | 40674922 | 40686447 | 90417 | KNSTRN | 3.03E-38 | NA | No | 66 |
| ENSG00000128928 | IVD | 15 | 40697686 | 40728146 | 3712 | IVD | NA | NA | Yes | 66 |
| ENSG00000128891 | C15orf57 | 15 | 40820882 | 40857256 | 90416 | C15orf57 | NA | NA | Yes | 66 |
| ENSG00000137824 | RMDN3 | 15 | 41028082 | 41048049 | 55177 | RMDN3 | NA | NA | Yes | 66 |
| ENSG00000104142 | VPS18 | 15 | 41186628 | 41196173 | 57617 | VPS18 | NA | NA | Yes | 66 |
| ENSG00000128908 | INO80 | 15 | 41271078 | 41408552 | 54617 | INO80 | NA | NA | No | 66 |
| ENSG00000178997 | EXD1 | 15 | 41474923 | 41522941 | 161829 | EXD1 | NA | NA | Yes | 66 |
| ENSG00000187446 | CHP1 | 15 | 41523037 | 41574043 | 11261 | CHP1 | NA | NA | Yes | 66 |
| ENSG00000104147 | OIP5 | 15 | 41601466 | 41624819 | 11339 | OIP5 | 0.0116217 | NA | Yes | 66 |
| ENSG00000137804 | NUSAP1 | 15 | 41624892 | 41673248 | 51203 | NUSAP1 | 2.52E-21 | + | Yes | 66 |
| ENSG00000137806 | NDUFAF1 | 15 | 41679551 | 41694717 | 51103 | NDUFAF1 | 1.75E-24 | - | No | 66 |
| ENSG00000067221 | STOML1 | 15 | 74275547 | 74286963 | 9399 | STOML1 | NA | NA | Yes | 67 |
| ENSG00000140464 | PML | 15 | 74287014 | 74340153 | 5371 | PML | NA | NA | Yes | 67 |
| ENSG00000138629 | UBL7 | 15 | 74738318 | 74753523 | 84993 | UBL7 | NA | NA | Yes | 67 |
| ENSG00000179361 | ARID3B | 15 | 74833518 | 74890472 | 10620 | ARID3B | NA | NA | Yes | 67 |
| ENSG00000179335 | CLK3 | 15 | 74890841 | 74932057 | 1198 | CLK3 | NA | NA | Yes | 67 |
| ENSG00000140465 | CYP1A1 | 15 | 75011883 | 75017951 | 1543 | CYP1A1 | NA | NA | Yes | 67 |
| ENSG00000103653 | CSK | 15 | 75074398 | 75095539 | 1445 | CSK | NA | NA | No | 67 |
| ENSG00000140506 | LMAN1L | 15 | 75105057 | 75118099 | 79748 | LMAN1L | NA | NA | Yes | 67 |
| ENSG00000213578 | CPLX3 | 15 | 75118888 | 75124141 | 594855 | CPLX3 | NA | NA | Yes | 67 |
| ENSG00000140474 | ULK3 | 15 | 75128457 | 75135687 | 25989 | ULK3 | NA | NA | Yes | 67 |
| ENSG00000140497 | SCAMP2 | 15 | 75136071 | 75165706 | 10066 | SCAMP2 | 4.15E-05 | + | Yes | 67 |
| ENSG00000178802 | MPI | 15 | 75182346 | 75191798 | 4351 | MPI | 2.47E-13 | + | No | 67 |
| ENSG00000178761 | FAM219B | 15 | 75192328 | 75199462 | 57184 | FAM219B | NA | NA | Yes | 67 |
| ENSG00000178741 | COX5A | 15 | 75212132 | 75230509 | 9377 | COX5A | NA | NA | Yes | 67 |
| ENSG00000178718 | RPP25 | 15 | 75246757 | 75249805 | 54913 | RPP25 | NA | NA | Yes | 67 |
| ENSG00000198794 | SCAMP5 | 15 | 75249560 | 75313837 | 192683 | SCAMP5 | NA | NA | Yes | 67 |
| ENSG00000167173 | C15orf39 | 15 | 75487984 | 75504510 | 56905 | C15orf39 | NA | NA | Yes | 67 |
| ENSG00000169371 | SNUPN | 15 | 75890424 | 75918810 | 10073 | SNUPN | NA | NA | Yes | 67 |
| ENSG00000102935 | ZNF423 | 16 | 49521435 | 49891830 | 23090 | ZNF423 | NA | NA | Yes | 68 |
| ENSG00000155393 | HEATR3 | 16 | 50099852 | 50140298 | 55027 | HEATR3 | NA | NA | Yes | 68 |
| ENSG00000141101 | NOB1 | 16 | 69775770 | 69788843 | 28987 | NOB1 | NA | NA | Yes | 69 |
| ENSG00000198373 | WWP2 | 16 | 69796209 | 69975644 | 11060 | WWP2 | 2.53E-10 | + | Yes | 69 |
| ENSG00000157322 | CLEC18A | 16 | 69984810 | 69998141 | 348174 | CLEC18A | 2.23E-29 | - | No | 69 |
| ENSG00000090857 | PDPR | 16 | 70147529 | 70195203 | 55066 | PDPR | 6.30E-06 | - | No | 69 |
| ENSG00000157335 | CLEC18C | 16 | 70207225 | 70221264 | 497190 | CLEC18C | 1.14E-05 | + | No | 69 |
| ENSG00000269746 | AC009060.1 | 16 | 70239303 | 70239683 | NA | NA | NA | NA | No | 69 |
| ENSG00000269866 | FKSG63 | 16 | 70258261 | 70258641 | NA | NA | NA | NA | No | 69 |
| ENSG00000223496 | EXOSC6 | 16 | 70284134 | 70285833 | 118460 | EXOSC6 | 3.91E-49 | - | No | 69 |
| ENSG00000090861 | AARS | 16 | 70286198 | 70323446 | 16 | AARS | 3.78E-09 | + | Yes | 69 |
| ENSG00000157349 | DDX19B | 16 | 70323566 | 70369186 | 11269 | DDX19B | NA | NA | Yes | 69 |
| ENSG00000260537 | RP11-529K1.3 | 16 | 70333097 | 70400163 | NA | NA | NA | NA | Yes | 69 |
| ENSG00000168872 | DDX19A | 16 | 70380732 | 70407286 | 55308 | DDX19A | NA | NA | No | 69 |
| ENSG00000157350 | ST3GAL2 | 16 | 70413338 | 70473140 | 6483 | ST3GAL2 | NA | NA | No | 69 |
| ENSG00000157353 | FUK | 16 | 70488324 | 70514177 | 197258 | FUK | 0.0342584 | + | No | 69 |
| ENSG00000103051 | COG4 | 16 | 70514471 | 70557468 | 25839 | COG4 | 0.00226371 | + | Yes | 69 |
| ENSG00000189091 | SF3B3 | 16 | 70557691 | 70608820 | 23450 | SF3B3 | 7.67E-12 | + | Yes | 69 |
| ENSG00000157368 | IL34 | 16 | 70613798 | 70694585 | 146433 | IL34 | 0.00075515 | + | No | 69 |
| ENSG00000132613 | MTSS1L | 16 | 70695107 | 70719969 | 92154 | MTSS1L | NA | NA | No | 69 |
| ENSG00000268927 | FLJ00418 | 16 | 70695570 | 70699739 | NA | NA | NA | NA | No | 69 |
| ENSG00000180917 | CMTR2 | 16 | 71315292 | 71323618 | 55783 | CMTR2 | 0.0123914 | NA | No | 69 |
| ENSG00000157429 | ZNF19 | 16 | 71498453 | 71598992 | 7567 | ZNF19 | NA | NA | Yes | 69 |
| ENSG00000166747 | AP1G1 | 16 | 71762913 | 71843104 | 164 | AP1G1 | NA | NA | Yes | 69 |
| ENSG00000176692 | FOXC2 | 16 | 86600857 | 86602539 | 2303 | FOXC2 | NA | NA | Yes | 70 |
| ENSG00000176678 | FOXL1 | 16 | 86609974 | 86615303 | 2300 | FOXL1 | NA | NA | Yes | 70 |
| ENSG00000260456 | C16orf95 | 16 | 87117168 | 87351022 | 100506581 | C16orf95 | NA | NA | No | 70 |
| ENSG00000007168 | PAFAH1B1 | 17 | 2496504 | 2588909 | 5048 | PAFAH1B1 | NA | NA | No | 71 |
| ENSG00000108599 | AKAP10 | 17 | 19807615 | 19881656 | 11216 | AKAP10 | NA | NA | Yes | 72 |
| ENSG00000128487 | SPECC1 | 17 | 19912657 | 20222339 | 92521 | SPECC1 | 3.13E-05 | NA | No | 72 |
| ENSG00000124422 | USP22 | 17 | 20902910 | 20947073 | 23326 | USP22 | NA | NA | Yes | 72 |
| ENSG00000154035 | C17orf103 | 17 | 21142183 | 21156722 | 256302 | C17orf103 | NA | NA | Yes | 72 |
| ENSG00000180329 | CCDC43 | 17 | 42750437 | 42767147 | 124808 | CCDC43 | NA | NA | Yes | 73 |
| ENSG00000108883 | EFTUD2 | 17 | 42927311 | 42977030 | 9343 | EFTUD2 | NA | NA | Yes | 73 |
| ENSG00000167131 | CCDC103 | 17 | 42976510 | 42982758 | 388389 | CCDC103 | NA | NA | Yes | 73 |
| ENSG00000214447 | FAM187A | 17 | 42977135 | 42982758 | 388389 | FAM187A | NA | NA | Yes | 73 |
| ENSG00000168517 | HEXIM2 | 17 | 43238067 | 43247407 | 124790 | HEXIM2 | NA | NA | Yes | 73 |
| ENSG00000184922 | FMNL1 | 17 | 43298811 | 43324687 | 752 | FMNL1 | 7.89E-19 | + | No | 73 |
| ENSG00000159314 | ARHGAP27 | 17 | 43471275 | 43511787 | 201176 | ARHGAP27 | 7.23E-14 | - | No | 73 |
| ENSG00000225190 | PLEKHM1 | 17 | 43513266 | 43568115 | 9842 | PLEKHM1 | 8.80E-35 | + | No | 73 |
| ENSG00000185294 | SPPL2C | 17 | 43922256 | 43924438 | 162540 | SPPL2C | 5.44E-19 | - | No | 73 |
| ENSG00000186868 | MAPT | 17 | 43971748 | 44105700 | 4137 | MAPT | NA | NA | Yes | 73 |
| ENSG00000256762 | STH | 17 | 44076616 | 44077060 | 246744 | STH | NA | NA | Yes | 73 |
| ENSG00000120071 | KANSL1 | 17 | 44107282 | 44302733 | 101929776 | KANSL1 | 1.59E-12 | + | No | 73 |
| ENSG00000228696 | ARL17B | 17 | 44352150 | 44439130 | 100506084 | ARL17B | 2.36E-17 | NA | No | 73 |
| ENSG00000176681 | LRRC37A | 17 | 44370099 | 44415160 | 474170 | LRRC37A | 4.87E-42 | - | No | 73 |
| ENSG00000238083 | LRRC37A2 | 17 | 44588877 | 44633016 | 474170 | LRRC37A2 | 7.91E-44 | - | No | 73 |
| ENSG00000185829 | ARL17A | 17 | 44594068 | 44657088 | 100506084 | ARL17A | 7.42E-31 | - | No | 73 |
| ENSG00000108379 | WNT3 | 17 | 44839872 | 44910520 | 101929777 | WNT3 | 2.06E-39 | - | Yes | 73 |
| ENSG00000158955 | WNT9B | 17 | 44910567 | 44964096 | 7484 | WNT9B | NA | NA | Yes | 73 |
| ENSG00000198336 | MYL4 | 17 | 45277812 | 45301045 | 4635 | MYL4 | NA | NA | Yes | 73 |
| ENSG00000168646 | AXIN2 | 17 | 63524681 | 63557765 | 8313 | AXIN2 | NA | NA | Yes | 74 |
| ENSG00000154240 | CEP112 | 17 | 63631656 | 64188202 | 201134 | CEP112 | NA | NA | No | 74 |
| ENSG00000091583 | APOH | 17 | 64208151 | 64252643 | 350 | APOH | NA | NA | No | 74 |
| ENSG00000154229 | PRKCA | 17 | 64298754 | 64806861 | 5578 | PRKCA | NA | NA | Yes | 74 |
| ENSG00000158270 | COLEC12 | 18 | 319361 | 500722 | 81035 | COLEC12 | NA | NA | Yes | 75 |
| ENSG00000197360 | ZNF98 | 19 | 22573821 | 22715287 | 148198 | ZNF98 | NA | NA | Yes | 76 |
| ENSG00000196172 | ZNF681 | 19 | 23921997 | 23941693 | 148213 | ZNF681 | 7.69E-14 | + | No | 76 |
| ENSG00000205246 | RPSAP58 | 19 | 23945807 | 24010937 | NA | RPSAP58 | NA | NA | No | 76 |
| ENSG00000213967 | ZNF726 | 19 | 24097678 | 24127961 | 730087 | ZNF726 | NA | NA | Yes | 76 |
| ENSG00000125740 | FOSB | 19 | 45971253 | 45978437 | 2354 | FOSB | NA | NA | Yes | 77 |
| ENSG00000125741 | OPA3 | 19 | 46030685 | 46105470 | 80207 | OPA3 | NA | NA | No | 77 |
| ENSG00000177464 | GPR4 | 19 | 46093022 | 46105466 | 2828 | GPR4 | NA | NA | No | 77 |
| ENSG00000125746 | EML2 | 19 | 46110252 | 46148887 | 24139 | EML2 | NA | NA | Yes | 77 |
| ENSG00000267757 | C19orf83 | 19 | 46144752 | 46146098 | 100287177 | C19orf83 | NA | NA | Yes | 77 |
| ENSG00000125743 | SNRPD2 | 19 | 46190712 | 46195827 | 6633 | SNRPD2 | NA | NA | Yes | 77 |
| ENSG00000011478 | QPCTL | 19 | 46195741 | 46207247 | 54814 | QPCTL | NA | NA | Yes | 77 |
| ENSG00000177051 | FBXO46 | 19 | 46213887 | 46234162 | 23403 | FBXO46 | NA | NA | Yes | 77 |
| ENSG00000237452 | AC074212.3 | 19 | 46236509 | 46267792 | NA | NA | NA | NA | Yes | 77 |
| ENSG00000171940 | ZNF217 | 20 | 52183604 | 52226446 | 7764 | ZNF217 | NA | NA | Yes | 78 |
| ENSG00000189060 | H1F0 | 22 | 38201114 | 38203442 | 3005 | H1F0 | NA | NA | Yes | 79 |
| ENSG00000100116 | GCAT | 22 | 38203912 | 38213183 | 23464 | GCAT | NA | NA | Yes | 79 |
| ENSG00000100124 | ANKRD54 | 22 | 38226862 | 38245334 | 129138 | ANKRD54 | NA | NA | Yes | 79 |
| ENSG00000100129 | EIF3L | 22 | 38244875 | 38285414 | 51386 | EIF3L | NA | NA | Yes | 79 |
| ENSG00000100139 | MICALL1 | 22 | 38301664 | 38338829 | 85377 | MICALL1 | NA | NA | No | 79 |
| ENSG00000100142 | POLR2F | 22 | 38348614 | 38437922 | 5435 | POLR2F | NA | NA | No | 79 |
| ENSG00000100151 | PICK1 | 22 | 38452318 | 38471708 | 9463 | PICK1 | 1.67E-18 | NA | Yes | 79 |
| ENSG00000100156 | SLC16A8 | 22 | 38474141 | 38480100 | 23539 | SLC16A8 | NA | NA | Yes | 79 |
| ENSG00000128298 | BAIAP2L2 | 22 | 38480896 | 38506677 | 80115 | BAIAP2L2 | NA | NA | No | 79 |
| ENSG00000184381 | PLA2G6 | 22 | 38507502 | 38601697 | 8398 | PLA2G6 | 0.0113973 | NA | Yes | 79 |
| ENSG00000185022 | MAFF | 22 | 38597889 | 38612518 | 23764 | MAFF | NA | NA | Yes | 79 |
| ENSG00000198792 | TMEM184B | 22 | 38615298 | 38669040 | 25829 | TMEM184B | NA | NA | Yes | 79 |
| ENSG00000213923 | CSNK1E | 22 | 38686697 | 38794527 | 1454 | CSNK1E | NA | NA | Yes | 79 |
| ENSG00000100201 | DDX17 | 22 | 38879445 | 38903665 | 10521 | DDX17 | NA | NA | Yes | 79 |
| ENSG00000100206 | DMC1 | 22 | 38914954 | 38966291 | 11144 | DMC1 | NA | NA | Yes | 79 |
| ENSG00000100226 | GTPBP1 | 22 | 39101728 | 39134304 | 9567 | GTPBP1 | NA | NA | Yes | 79 |
| ENSG00000221890 | NPTXR | 22 | 39214457 | 39239987 | 23467 | NPTXR | NA | NA | Yes | 79 |
| ENSG00000244509 | APOBEC3C | 22 | 39410088 | 39416357 | 27350 | APOBEC3C | NA | NA | Yes | 79 |
| ENSG00000100321 | SYNGR1 | 22 | 39745930 | 39781593 | 9145 | SYNGR1 | NA | NA | Yes | 79 |
| ENSG00000100324 | TAB1 | 22 | 39795746 | 39833065 | 10454 | TAB1 | NA | NA | Yes | 79 |
| ENSG00000100335 | MIEF1 | 22 | 39895437 | 39914137 | 54471 | MIEF1 | NA | NA | Yes | 79 |
| ENSG00000186732 | MPPED1 | 22 | 43807202 | 43903728 | 758 | MPPED1 | 6.39E-06 | - | No | 80 |